\documentclass{jkas}

\def\year{2025} 
\def\volume{} 
\def\issue{1} 
\def\beginpage{1} 
\def\received{---} 
\def\accepted{---} 
\def\published{---} 
\date{Received \received; Accepted \accepted; Published \published}

\newcommand\ion[2]{{#1}\,{\sc #2}} 

\title{%
K-DRIFT Science Theme:\\
Illuminating the Next Era of Galaxy Cluster Science
}

\author[1,2]{Jaewon Yoo$\star$}{0000-0002-6841-8329}
\author[1]{Kyungwon Chun$\star$}{0000-0001-9544-7021}
\author[1,3]{Jongwan Ko}{0000-0002-9434-5936}
\author[1,3]{Jihye Shin}{0000-0001-5135-1693}
\author[4]{Cristiano G. Sabiu}{0000-0002-5513-5303}
\author[1,5]{Jaehyun Lee}{0000-0002-6810-1778}
\author[1,3]{Kwang-il Seon}{0000-0001-9561-8134}
\author[1]{Jae-Woo Kim}{0000-0002-1710-4442}
\author[1, 6]{Jinsu Rhee}{0000-0002-0184-9589}
\author[1]{Sungryong Hong}{0000-0001-9991-8222}
\author[1]{Woowon Byun}{0000-0002-7762-7712}
\author[7]{Hyowon Kim}{0000-0003-4032-8572}
\author[1]{Sang-Hyun Chun}{0000-0002-6154-7558}
\author[1,3]{Hong Soo Park}{0000-0002-3505-3036}
\author[1,8]{Yongmin Yoon}{0000-0003-0134-8968}
\author[1,3]{Jeehye Shin}{0000-0002-9914-3129} 

\affil[1]{Korea Astronomy and Space Science Institute, Daejeon 34055, Republic of Korea}
\affil[2]{Quantum Universe Center, Korea Institute for Advanced Study, Seoul 02455, Republic of Korea}
\affil[3]{Department of Astronomy and Space Science, University of Science and Technology, Korea, Daejeon 34113, Republic of Korea}
\affil[4]{Natural Science Research Institute, University of Seoul, Seoul 02504, Republic of Korea}
\affil[5]{School of Physics, Korea Institute for Advanced Study, Seoul 02455, Republic of Korea}
\affil[6]{Institut d’Astrophysique de Paris, Sorbonne Université, CNRS, UMR 7095, 98 bis bd Arago, 75014 Paris, France}
\affil[7]{Departamento de F{\'i}sica, Universidad T{\'e}cnica Federico Santa Mar{\'i}a, Avenida Vicu{\~n}a Mackenna 3939, San Joaqu{\'i}n, Santiago, Chile}
\affil[8]{Department of Astronomy and Atmospheric Sciences, Kyungpook National University, Daegu 41566, Republic of Korea}

\def\corrauthor{%
J. Yoo, \email{jwyoo@kasi.re.kr}
K. Chun, \email{kwchun@kasi.re.kr}
}

\def\runningauthor{%
Yoo et al.
}

\def\runningtitle{%
K-DRIFT: Galaxy Cluster
}

\def\keywords{%
galaxies: general --- galaxies: interactions --- galaxies: stellar content --- galaxies: structure --- galaxies: dwarf galaxies --- galaxies: low surface brightness galaxies
}

\def\abstracttext{%
The KASI Deep Rolling Imaging Fast Telescope (K-DRIFT) is a pioneering instrument designed to explore low-surface-brightness (LSB) phenomena. This white paper presents a compelling array of science cases that showcase K-DRIFT's unique capabilities in unraveling the mysteries of intracluster light (ICL) and other LSB components within galaxy clusters.
Exploring the origin of ICL in galaxy clusters and comparing the spatial distributions of ICL and dark matter will offer new insights into galaxy cluster dynamics.
Moreover, investigating LSB objects in galaxy clusters, such as LSB structures in the brightest cluster galaxy, ultra-diffuse galaxies, and tidal features, will enhance our understanding of galaxy evolution within the cluster environment.
We present our strategies for addressing scientific queries, encompassing LSB observation and analysis techniques, specialized simulations, and machine-learning approaches.
Additionally, we examine the potential synergies between K-DRIFT and other ongoing or forthcoming multi-wavelength surveys. 
This white paper advocates for the recognition and support of K-DRIFT as a dedicated tool for advancing our understanding of the universe's subtlest phenomena.
}

\begin{document}
\jkashead 


\section{Introduction}

Galaxy clusters, the largest gravitationally bound structures in the observable universe, are pivotal to our understanding of the cosmos. Encompassing a rich tapestry of galaxies, dark matter, and intergalactic gas, these clusters offer an unparalleled window into the underlying forces and processes that shape the universe. Their abundance, spatial distribution, and internal dynamics are sensitive to cosmological parameters such as matter density, the amplitude of density fluctuations, and the nature of dark matter and dark energy. The mass function derived from galaxy cluster statistics constrains theoretical models, while dramatic systems like the Bullet Cluster provide direct evidence for dark matter through gravitational lensing and the separation from ordinary matter during high-speed collisions \citep{2004ApJ...604..596C}.

Formed hierarchically through the mergers of smaller structures—galaxies, groups, and filaments—galaxy clusters represent the final stage of structure formation. This assembly process drives the co-evolution of baryons and dark matter and shapes the dynamical state of clusters. Residing in the densest regions of the universe, clusters provide an extreme environment for studying galaxy interactions, hot X-ray emitting gas, and the transformation of galaxies in high-density regions. The typical cluster mass composition—approximately 85\% dark matter, 12\% gas, and 3\% stars \citep{2009ApJ...693.1142S, 2013ApJ...778...14G, 2013MNRAS.429.3288S}—underscores the dominant role of dark matter in structuring the universe, with weak gravitational lensing serving as the most direct method for mapping its distribution.

A rich galaxy cluster often contains a well-developed brightest cluster galaxy (BCG), and the surrounding intracluster light (ICL; the diffuse light from stars not gravitationally bound to any individual cluster galaxy) traces the cumulative effects of mergers and accretion events predicted in $\Lambda$CDM models. The ICL, which is collisionless and dynamically associated with the global cluster potential, is emerging as a promising visible tracer for dark matter \citep{2019MNRAS.482.2838M, 2020MNRAS.494.1859A, 2022ApJS..261...28Y, 2024ApJ...965..145Y, 2025ApJ...988..229Y}. Additionally, $\Lambda$CDM simulations predict a rich population of low-surface-brightness (LSB) structures, including dwarf galaxies and tidal features, many of which remain undetected in current surveys, contributing to longstanding discrepancies such as the \textit{missing satellite problem} \citep{1999ApJ...522...82K}. Studying ICL and LSB features can thus offer crucial clues about hierarchical structure formation and the distribution of dark matter \citep{2005ApJ...631L..41M, 2016IAUS..317...27M}.

Numerical simulations are a powerful tool for understanding the observed properties of galaxy clusters and their formation and evolution.
In general, cosmological hydrodynamic simulations have been used for this purpose, as they allow us to trace the formation and evolution of stellar structures self-consistently, with stars forming from the gas component whose hydrodynamics is modeled, e.g.,The Three Hundred project \citep{cui2018}, IllustrisTNG \citep{2018MNRAS.475..624N}, ROMULUSC \citep{tremmel2019}, and Horizon Run 5 \citep[HR5;][]{lee21}.
However, modeling the baryonic components is highly computationally expensive, which limits either the number of clusters that can be simulated or the achievable mass or spatial resolution.

To overcome these limitations, some studies have employed an alternative technique.
These simulations focus on tracing the structures in the LSB universe much faster and at higher resolution than cosmic hydrodynamic simulations \citep[e.g.,][]{2006ApJ...648..936R,harris2017,chun2022}.
Although they lack hydrodynamic processes and in-situ star formation, they complement full-physics simulations and have demonstrated that LSB structures are essential for understanding the observed properties of clusters and their formation and evolution \citep[e.g.,][]{2020MNRAS.494.1859A,2021MNRAS.501.1300S,2021MNRAS.508.2634Y,chun2023,tang2023,contreras-santos2024,chun2024,2024ApJ...965..145Y}.
It is evident, therefore, that observations of the LSB universe would provide valuable data to test and constrain models of galaxy and galaxy cluster evolution as well as cosmology. 

The forthcoming KASI Deep Rolling Imaging Fast Telescope (K-DRIFT) stands poised to revolutionize our understanding of these phenomena. With its advanced capabilities for observing LSB phenomena, K-DRIFT will illuminate the faint and elusive components of galaxy clusters, offering unprecedented insights into their formation, evolution, and the dark matter that pervades them. The synergy between K-DRIFT and other large-scale, multi-wavelength surveys will further enhance our comprehension of the universe, bridging gaps in current knowledge and paving the way for new discoveries.

This paper is organized as follows. In Section \ref{sec:science}, we describe the general science cases that K-DRIFT will address. Section \ref{sec:strategy} outlines our strategies for survey design, simulations, and machine-learning approaches tailored to specific scientific goals. Section \ref{sec:prospects} discusses the potential synergies between K-DRIFT and other major ongoing or future surveys. In Section \ref{sec:conclusions}, we conclude by summarizing the implications of K-DRIFT's contributions to astrophysics and cosmology.

\section{Science Cases for K-DRIFT \label{sec:science}}
\begin{figure*}
\centering
\includegraphics[width=2\columnwidth]{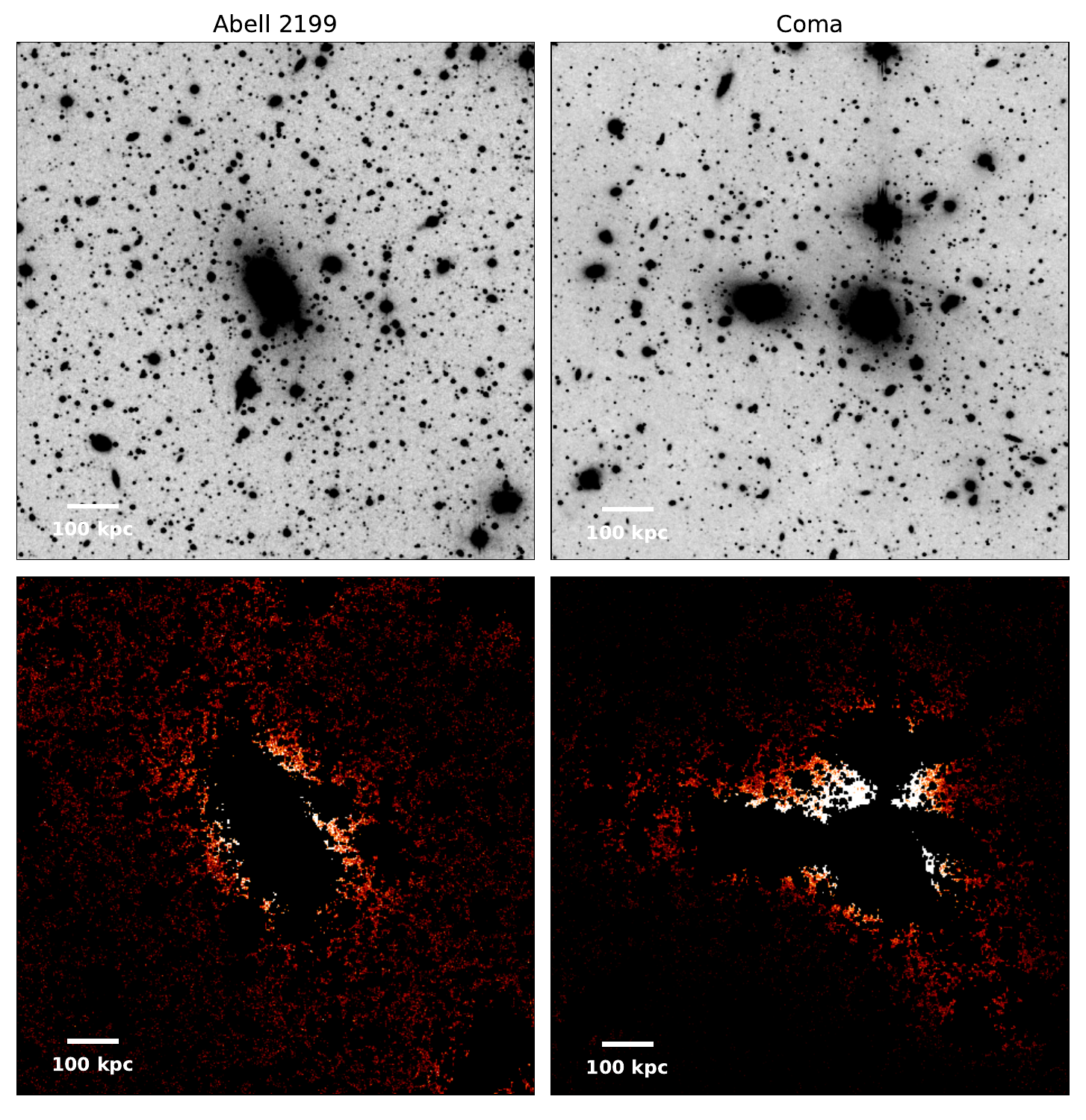}
\caption{Galaxy cluster images observed with the K-DRIFT Pathfinder. Left: Abell 2199, spanning approximately $25 \times 25$ arcmin$^2$. Right: Coma cluster, spanning approximately $32 \times 32$ arcmin$^2$. In the upper panels, regions with surface brightness brighter than 26.5 mag arcsec$^{-2}$ (the typical ICL detection threshold) are masked and displayed in the lower panels. The K-DRIFT G1 is expected to provide wider FoV and improved image quality.}
\label{fig:clusters}
\end{figure*}

The K-DRIFT Pathfinder at the Bohyunsan Optical Astronomy Observatory in Korea has already demonstrated a surface-brightness sensitivity of $\mu_{r,1\sigma} \sim$ 28.5 mag arcsec$^{-2}$ \citep{2022PASP..134h4101B}. Figure \ref{fig:clusters} shows the Abell 2199 and Coma clusters observed by the K-DRIFT Pathfinder.
The next-generation K-DRIFT, `K-DRIFT Generation 1 (G1)', will represent a major advancement in the study of LSB phenomena ($\mu_V >$ 29 mag arcsec$^{-2}$). With its wide field-of-view (FoV) of approximately 4.5 $\times$ 4.5 deg$^2$ and a pixel scale of about 2~arcsec, K-DRIFT G1 is ideally suited for expansive and detailed night sky surveys. The expected depth of K-DRIFT G1, reaching 
$\mu_{r}^{\mathrm{limit}}$(3$\sigma$, $10^{\prime\prime} \times 10^{\prime\prime}$) $\sim$ 29-30 mag arcsec$^{-2}$
in 10 hours of observation, will mark it a powerful tool for probing the faintest structures in the universe. 
A detailed description of the telescope and survey strategy is provided in \cite{2025arXiv251022250K}.
This section explores the LSB science cases that can be investigated using K-DRIFT G1, with particular focus on the ICL within galaxy clusters.

\subsection{Origin of Intracluster Light}


The phenomenon of ICL has garnered significant interest in recent years, thanks to deep observations of galaxy clusters that have revealed distinct diffuse light not bound to individual galaxies \citep[][see Figure 5 in \citealp{2018ApJ...862...95K}]
{1951PASP...63...61Z, 1998Natur.396..549G, 2002ApJ...575..779F, 2004ApJ...617..879L, 2005ApJ...618..195G, 2005MNRAS.358..949Z, 2005ApJ...631L..41M, 2017ApJ...834...16M, 2018MNRAS.474.3009D, 2019A&A...622A.183J, 2020ApJS..247...43K, 2021MNRAS.502.2419F, 2021ApJ...910...45M, 2021MNRAS.508.2634Y}. 

The study of ICL with K-DRIFT can significantly enhance our understanding of the baryonic component of the universe. By quantifying the extent and characteristics of ICL, we may address the current discrepancy between the observed and simulated dark matter and baryon fractions in galaxy clusters \citep{2016ApJ...826..146B}. Observations of the ICL distribution and its correlation with other cluster properties (such as total mass, galaxy density, and dynamical state) are expected to place strong constraints on theoretical models of galaxy cluster evolution \citep{2004ApJ...617..879L, 2010HiA....15...97A}. Furthermore, the detailed mapping of ICL can provide insights into the processes of galaxy interactions and mergers, offering a deeper understanding of the mechanisms driving the evolution of clusters.

Although there is growing consensus on the importance of ICL study, its main formation mechanism remains under debate due to the limited number of ICL measurements across a wide range of redshifts and masses, as well as the disparity in observational methods of measuring the ICL and its properties. The ICL origins suggested in previous observation and simulation studies are as follows: major mergers of BCGs and subsequent violent relaxation \citep{2007ApJ...665L...9R, 2018ApJ...862...95K},
tidal stripping from the outskirts of L$^{\ast}$ member galaxies \citep{2017ApJ...851...75I, 2018MNRAS.474.3009D, 2018MNRAS.474..917M},
disruptions of dwarf galaxies as they fall toward galaxy cluster centers \citep{2007ApJ...666...20P, 2007MNRAS.377....2M, 2007ApJ...668..826C, 2011MNRAS.414..602T}, 
in-situ star formation \citep{2002ApJ...580L.121G,2010MNRAS.406..936P,2022ApJ...930...25B} (however, several observational results disagree with this scenario \citep{2011ApJ...729..142S,2012MNRAS.427..850M}), and
\textit{pre-processing} in galaxy groups, where intragroup light formed in the infalling systems later becomes part of the ICL in the main cluster \citep{2004PASJ...56...29F,2004ApJ...617..879L, 2004cgpc.symp..277M, 2005MNRAS.357..478S, 2006ApJ...648..936R,2014MNRAS.437.3787C}.

The morphological signatures of ICL can vary depending on its origin. If formed predominantly through recent tidal stripping or galaxy disruption, the ICL may appear as distinct tidal features, such as streams or shells, that preserve the imprint of those interactions. In contrast, ICL originating from early-phase BCG major mergers is expected to be more smoothly distributed and diffuse, reflecting the violent relaxation processes associated with such events.

The color and metallicity profiles of ICL can also provide clues to its origin. A negative gradient in these profiles suggests disruption and tidal stripping as primary mechanisms, whereas a near-constant gradient could imply major mergers of massive galaxies. The composition of the ICL, whether dominated by metal-poor stars from low-mass satellite stripping or metal-rich stars from massive galaxy mergers, offers further insights into its formation history. The age of the ICL, whether predominantly old or young, can shed light on its origin, reflecting contributions from the stripping of old stellar systems or younger, star-forming galaxies and in-situ formation processes.

With its deep imaging capabilities in multiple bands, K-DRIFT is uniquely positioned to advance our understanding of ICL origins. By measuring the morphology, color, and color gradients of ICL and comparing them with the overall light of galaxy clusters, K-DRIFT can provide critical data to constrain the possible origins of ICL in various clusters. Its ability to observe faint and diffuse light with high sensitivity makes it an ideal instrument for disentangling the complex processes that contribute to the formation of ICL, thereby shedding light on the dynamical evolution and assembly history of galaxy clusters.

\subsection{Spatial Distribution of ICL and Dark Matter}\label{subsec:DMtracer}
{\bf ICL as a dark matter tracer: }
Weak gravitational lensing is a powerful method to trace dark matter. However, it has several drawbacks, such as the difficulty of identifying background galaxies and accurately measuring their shapes in the presence of the telescope point spread function (PSF). These issues all contribute to an increase in the uncertainty of the derived maps. It would be advantageous to search for other possible dark matter tracers that are simpler to observe and/or are subject to different systematic biases.
The ICL is collisionless and expected to follow the global potential well of the galaxy cluster. Therefore, it could serve as a luminous tracer for dark matter \citep{2018MNRAS.474..917M, 2019MNRAS.482.2838M, 2020MNRAS.494.1859A, 2022ApJS..261...28Y, 2024ApJ...965..145Y, 2025ApJ...988..229Y}. If the spatial distribution of the ICL coincides with that of the dark matter, it may be possible to construct a dark matter map without relying on laborious weak lensing analyses.

\begin{figure}
\centering
\includegraphics[width=84mm,trim={0 0 1.5cm 1.1cm},clip]{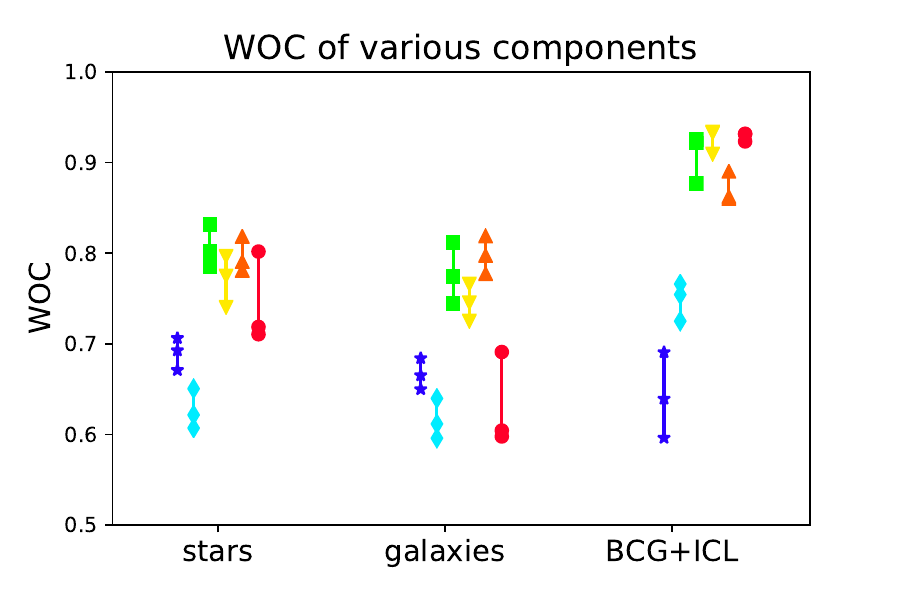}
\includegraphics[width=84mm,trim={0 0 1.5cm 1.1cm},clip]{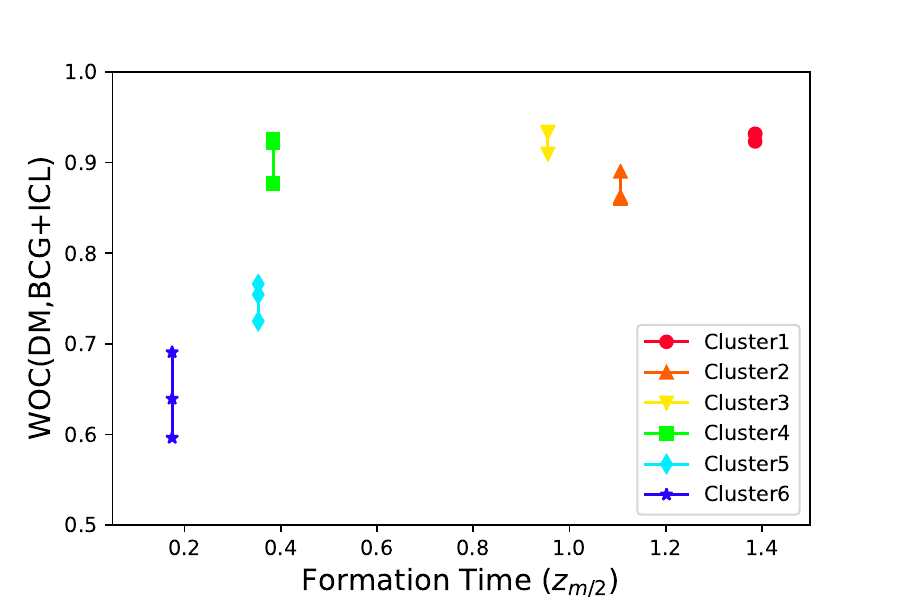}
\caption{  
Top: the WOC results for different components of galaxy clusters.  
The three data points with the same symbols come from the three projection angles of each galaxy cluster, which are connected with lines. Bottom: the WOC results for different dynamical state of galaxy clusters. Figure credit from \cite{2022ApJS..261...28Y}.
\label{fig:Component}}
\end{figure}

\cite{2022ApJS..261...28Y} introduced the \textit{Weighted Overlap Coefficient} (WOC) method to quantify the similarity between the spatial distributions of dark matter and ICL. Their findings indicate that the combination of ICL with the BCG, exhibiting the highest WOC value among various cluster components, can serve as an effective luminous tracer for dark matter in galaxy clusters (see the upper panel of Figure \ref{fig:Component}).

The forthcoming galaxy cluster survey conducted by K-DRIFT promises to deliver deep imaging data that is well suited for constructing comprehensive two-dimensional maps of the ICL. Harnessing the synergy between this imaging data and weak-lensing measurements will enable us to explore the feasibility of employing the ICL as a tracer for dark matter, therby providing compelling observational evidence supporting this prospect.

{\bf Constraining dark matter models using the ICL: } 
The self-interacting dark matter (SIDM) model has been attracting increasing attention. First, there is no a priori reason why dark matter particles should not interact with each other \citep{2000PhRvL..84.3760S, 2000ApJ...534L.143B}. Weak self-interactions are a natural consequence of certain particle physics theories concerning the origin of dark matter \citep{2018PhR...730....1T}. Moreover, introducing the SIDM could help resolve tensions between the results of dark matter-only simulations and observations of dwarf and low-mass galaxies \citep{2017ARA&A..55..343B}.


The Cluster-EAGLE zoom-in cosmological simulation \citep{2017MNRAS.470.4186B} contains 30 galaxy clusters ($>10^{14} M_{\odot}$) in a $\Lambda$CDM universe. Two of these clusters have been re-simulated from identical initial conditions in a $\Lambda$SIDM universe \citep{2018MNRAS.476L..20R}.
The SIDM was assumed to have an isotropic, velocity-independent interaction cross-section of $\sigma/m$ = 1 cm$^2$ g$^{-1}$. During each simulation timestep, $\Delta t$, dark matter particles scatter elastically off neighbors within a radius $h_\mathrm{SI}$ = 2.66 kpc with a probability
$$P_\mathrm{scat}=\frac{(\sigma/m)m_\mathrm{DM}\nu\Delta t}{\frac{4\pi}{3}h^3_\mathrm{SI}}$$
where $\nu$ is the particle's relative velocity and $m_\mathrm{DM}$ the dark matter particle mass \citep{2017MNRAS.467.4719R}.

In the SIDM universe, when a galaxy falls into a SIDM cluster, interactions between its dark matter particles and those of the cluster could scatter dark matter out of the galaxy. This \textit{evaporation} process acts in addition to tidal stripping, accelerating the overall mass loss. Consequently, the SIDM model would predict different tidal interaction histories from the cold dark matter (CDM) model \citep{2022MNRAS.511.5927S}. Indeed, the simulated galaxy clusters in the SIDM universe showed fewer survived galaxies and more disrupted galaxies than those in the CDM universe with the same initial conditions (see Figure 7 in \citealp{2022MNRAS.511.5927S}).

Variations in tidal interaction histories, such as stronger tidal stripping from galaxy outskirts and an increased number of disrupted galaxies, would lead to corresponding differences in both the amount and morphology of ICL within galaxy clusters.
Therefore, comparing the spatial distribution similarity between dark matter and the ICL could help us constrain dark matter models such as the SIDM or the CDM.

\subsection{ICL and Dynamical Evolution of Galaxy Clusters}
According to the standard model of structure formation, galaxy clusters are formed through successive mergers and accretion of smaller systems provided along filament structures in the large-scale structure \citep{1970A&A.....5...84Z}. These dynamical processes leave signatures in the cluster properties before the system becomes fully virialized, such as asymmetric, multiple peaks in the X-ray \citep{2006A&A...447..827K} and/or galaxy density field \citep{2009A&A...495...15B}, which imply the existence of substructures. The system will eventually achieve virialization and relaxation following a dynamic young phase characterized by active merging events. 

In galaxy cluster simulations, attempts to parametrize the dynamical state of a galaxy cluster involves several quantities including the center-of-mass offset (distance between the center of mass and the density peak of the system), the subhalo mass fraction, the virial ratio ($\eta$ = 2$T/|W|$, where $T$ and $W$ are the kinetic and the gravitational potential energies, respectively), the redshift of the last major merger, and the redshift at which the cluster reaches 50\% of its present-day mass ($z=0$). In observations, on the other hand, commonly used indicators include the offset between the BCG and the X-ray peak \citep{2003ApJ...585..687K}, the magnitude gap between the BCG and the second brightest galaxy \citep{2003MNRAS.343..627J}, and the substructure luminosity fraction \citep{2013MNRAS.436..275W}.

{\bf  BCG+ICL fraction as an indicator of galaxy cluster dynamical state:} 
The abundance of ICL, together with the growth of the BCG, increases through the dynamical exchange of galaxies within clusters, as shown in numerous simulation studies \citep{2007MNRAS.377....2M, 2007ApJ...666...20P, 2007ApJ...668..826C, 2010MNRAS.406..936P, 2011ApJ...732...48R, 2014MNRAS.437.3787C, 2015MNRAS.451.2703C}. Thus, the buildup of ICL reflects the cumulative effect of galaxy interactions during cluster evolution. Within this framework, the combined fraction of BCG and ICL light relative to the total cluster light (hereafter the \textit{BCG+ICL fraction}) can serve as a valuable proxy for assessing the dynamical state of galaxy clusters. More evolved systems are expected to display higher BCG+ICL fractions than dynamically younger clusters, consistent with both observational and simulation results \citep{2006ApJ...648..936R, 2006A&A...457..771A, 2008MNRAS.388.1433D, 2018MNRAS.474..917M, chun2023, 2024ApJ...965..145Y}. 

However, measurements of ICL fractions reported in the literature vary widely, ranging from as low as 2.6\% \citep{2010MNRAS.403L..79M} to as high as 50\% \citep{2004ApJ...617..879L}. This large spread arises from differences in the dynamical states of the observed clusters, the adopted definitions of ICL, and systematic uncertainties associated with telescope instrumentation and data reduction methods (see Figure 15 in  \citealp{2021MNRAS.508.2634Y}). 

The upcoming K-DRIFT cluster survey holds promise in establishing the connection between the BCG+ICL fraction and the dynamical state of galaxy clusters. This potential will be realized through a consistent ICL definition and uniform observations facilitated by a single instrument.


{\bf Spatial distribution similarity with dark matter:} More relaxed galaxy clusters, which have had enough time to virialize, could show stronger alignment between the distributions of dark matter and other cluster components. If the BCG+ICL component of more relaxed galaxy clusters exhibits a spatial distribution more similar to that of the dark matter compared to dynamically younger galaxy clusters, then this similarity could be used as an indicator of the cluster's dynamical state. 

Accordingly, we investigated how the two-dimensional spatial distribution similarity between dark matter and BCG+ICL varies with the formation time ($z_{m/2}$; the redshift at which the cluster has accumulated half of its final observed mass), a proxy for the relaxation level of galaxy clusters \citep{2022ApJS..261...28Y, 2024ApJ...965..145Y}. Among the sample galaxy clusters, the more relaxed clusters showed a stronger similarity between the spatial distribution of the dark matter and BCG+ICL than the dynamically young clusters (see the lower panel in Figure \ref{fig:Component}). The relationship between the dark matter—BCG+ICL similarity and $z_{m/2}$ shows a clear trend, although cluster 4 appears to be an outlier. Among the clusters with high similarity (clusters 1, 2, 3, and 4), cluster 4, which has the latest formation time and is therefore the most dynamically unrelaxed, exhibits slightly larger variations with viewing angle.

The upcoming K-DRIFT deep imaging data, uniformly surveyed across galaxy clusters, will provide a valuable opportunity to extensively study the relationship between the spatial distribution similarity of ICL and dark matter, and the dynamical state of the host galaxy cluster.



\subsection{Brightest and Faintest Ends}

{\bf BCG formation:} The BCG is the most luminous and massive galaxy in a cluster. It is generally located near the cluster center and can be more than two magnitudes brighter than the second brightest galaxy \citep[e.g.,][]{raouf2019}. 

The coalescence of infalling galaxies is thought to be the major formation mechanism of BCGs~\citep[e.g.,][]{1977ApJ...217L.125O,1983ApJ...268...30R,2007MNRAS.375....2D}. Because they mostly reside at the centers or in the vicinity of the deepest gravitational potential wells of clusters, BCGs can accrete baryons far more efficiently than other cluster galaxies. This leads to larger characteristic sizes, higher dark matter content, and greater velocity dispersions compared to typical elliptical galaxies \citep[e.g.,][]{2007MNRAS.379..867V,2015MNRAS.453.4444Z,2020ApJS..247...43K}.

According to most observational and theoretical studies, a significant portion of stars in BCGs likely formed {\it ex-situ} at $z>2$ and were later accreted into BCGs via mergers \citep[e.g.,][]{2007MNRAS.375....2D,2013ApJ...766...38L,2016MNRAS.458.2371R,2017ApJ...836..161L, 2018MNRAS.479..932C,2019ApJ...881..150C}. However, the timescale of this build-up process is still debatable \citep[see][and references therein]{chu2021}.
Some studies have suggested that the sizes of BCGs do not evolve significantly after $z\sim1$ \citep{2011MNRAS.414..445S,2017MNRAS.465.2101O}, whereas others have shown that BCGs have more than doubled in size relatively recently ~\citep[$< 6~$ Gyr;][]{2015ApJ...802...73S}. It has also been suggested that BCGs have almost halted the growth of their inner structures since $z\sim0.4$, while they have predominantly built up outside structures, including ICL, ever since~\citep{2020MNRAS.491.3751D}. This may result in two-S\'{e}ric component galaxies with shallower luminosity profiles in their outer regions. \citet{2022A&A...666A..54C} proposed that the two-component nature of BCGs—a compact inner component and a shallower extended outer envelope—can be resolved in deep surveys reaching surface brightness levels fainter than 26~mag~arcsec$^{-2}$. Given that K-DRIFT aims to reach a depth of $\sim$30~mag~arcsec$^{-2}$, it will enable us to distinguish the outer component from the inner structure, thereby offering a crucial test for BCG formation scenarios.

{\bf Faint-end slope of the luminosity function:} \citet{kim23} demonstrates, using the HR5 simulation~\citep{lee21}, that the surface-brightness and aperture limits of observations substantially affect the galaxy stellar mass function at both the low- and high-mass ends. Since BCGs typically have extended luminosity profiles, their total luminosity or mass can be underestimated in surveys with shallow depth or small apertures. At the low-mass end, observations inevitably miss a large fraction of LSB galaxies when the surface-brightness limit is insufficient. Thanks to its unprecedented depth and wide FoV, K-DRIFT will enable us to efficiently capture the LSB components of the BCGs and find LSB galaxies in clusters. 

\subsection{Environmental Effects on Galaxies}

In the concordance $\Lambda$CDM cosmology, initial density peaks gradually accumulate matter spread over the radius of tens of cMpc, eventually forming galaxy clusters. The inner environments of galaxy clusters significantly influence infalling galaxies through their high gas densities and strong tidal fields. This inevitably makes cluster satellites differ from those in field environments.

Environmental effects are strongly exerted on cluster galaxies through several different physical processes. High-speed tidal encounters between galaxies occur more frequently in clusters than in the field, resulting in mass loss and morphological transformation of the cluster satellites~\citep[Harassment;][]{1996Natur.379..613M}. Tidal stripping of halos truncates the hot gas envelope of galaxies, the outer source of the interstellar medium (ISM), eventually suppressing star formation \citep[Strangulation;][]{2015Natur.521..192P}. As cluster satellite galaxies move through the intracluster medium (ICM), they experience ram pressure that can directly strip the ISM from galaxies~\citep[Ram pressure stripping (RPS);][]{1972ApJ...176....1G}. This also leads to the quenching of star formation in cluster galaxies~\citep{koopmann04a,koopmann04b}.

Environmental effects sometimes leave characteristic features in cluster satellite galaxies. For example, ram pressure stripping can form ``jellyfish galaxies'' \citep[e.g.,][]{lee20,lee22,2022ApJ...931L..22L}, and galaxies that have experienced harassment often develop stellar streams \citep[e.g.,][]{2010MNRAS.405.1723S,2015MNRAS.454.2502S}.
The presence of these structures can provide valuable information about the evolutionary path of galaxies in the cluster environment \citep[e.g.,][]{2015MNRAS.454.2502S,2019MNRAS.483.1042Y,2021ApJ...912..149S,2022ApJ...934...86S}, but such structures typically have very low surface brightness.
Therefore, we expect that K-DRIFT observations will reveal evidence of morphological transformation of member galaxies and enhance our understanding of their evolutionary paths.

{\bf UDGs in the cluster:} Ultra-diffuse galaxies (UDGs) are galaxies with an extremely low surface brightness \citep[$\mu_g >$ 24 mag arcsec$^{-2}$;][]{2015ApJ...798L..45V,2015ApJ...807L...2K}. They have an effective radius comparable to that of large galaxies ($r_{\rm{eff}} >$ 1.5 kpc), but their stellar mass is similar to that of dwarf galaxies ($M_{\star} = 10^{7-8} M_{\odot}$).

After \citet{2015ApJ...798L..45V} discovered 47 LSB galaxies in the Coma cluster and proposed the term UDGs for these galaxies, several studies have identified UDGs in other clusters and in the field \citep{2015ApJ...809L..21M,2016ApJ...833..168M,2017ApJ...836..191T,2017ApJ...846...26S,2020ApJ...899...69L}. Although UDGs generally have low luminosities and large sizes, their stellar populations exhibit variations. In particular, UDGs in the clusters show a diverse range of properties, even if they are located within the same cluster. Observational and theoretical studies have indicated that the evolution of UDGs within cluster environments is the main driver of their diverse properties.

Because UDGs have diverse properties, various scenarios have been suggested to explain their formation \citep[e.g.,][]{2015MNRAS.452..937Y,2016MNRAS.459L..51A,2017MNRAS.466L...1D,sales2020,2020ApJ...894...75L}. In particular, the unique conditions of the cluster environment make cluster satellites evolve into UDGs as they undergo tidal stripping. \citet{sales2020} showed that tidal-origin UDGs have distinct characteristics compared to those formed by internal processes in the field. Specifically, these tidal-origin UDGs exhibit lower velocity dispersion, higher metallicities, and lower dark matter content. 
Furthermore, the distribution of tidal-origin UDGs peaks at the center of clusters, which is in contrast to previous observations suggesting that UDGs are rare at cluster centers.

Contrary to previous observations, \citet{2020ApJ...899...69L} found that the distribution of 44 UDGs identified by the Next Generation Virgo Cluster Survey (NGVS) is more concentrated than other galaxies with similar luminosity. Additionally, due to their low surface brightness, half of the 44 UDGs found by NGVS had not been detected in earlier observational studies. These discrepancies may be attributed to the challenges associated with UDG observations, especially within the cluster's central regions. Consequently, conducting deeper imaging observations using K-DRIFT for other clusters may provide insight into whether the distribution of UDGs in that cluster is a unique attribute of the Virgo cluster or a natural result of UDG formation. 

{\bf Tidal featured galaxies:} In cluster environments, satellite galaxies experience the strong tidal field of the galaxy cluster or interactions with other galaxies. 
In this process, galaxies lose mass, and tidal features can be formed.
Moreover, since about 50\% of satellite galaxies have experienced preprocessing \citep[e.g.,][]{han2018}, some satellites form tidal features even before falling into the cluster.

\cite{adams2012} analyzed 54 galaxy clusters at $0.04 < z < 0.15$ containing 3551 early-type galaxies and found that $\sim$3\% of cluster early-type galaxies have tidal features of $\mu_{r'}<26.5~\rm{mag}~\rm{arcsec^{-2}}$.
They also showed that there is a decrease in the fraction of galaxies with tidal features at smaller clustercentric radii.
On the other hand, \cite{sheen2012} showed that there is no trend in the fraction of galaxies with merging features of $\mu_{r'}<28~\rm{mag}~\rm{arcsec^{-2}}$ depending on the clustercentric radius.
Furthermore, they found post-merger features in 24\% of red-sequence galaxies ($M_{r} < -20$ mag) in four rich Abell clusters at $z < 0.1$.
In a more recent study, \cite{oh2018} carried out the KASI-Yonsei Deep Imaging Survey of Clusters, targeting 14 Abell clusters at $0.015 < z < 0.144$ and, found that 20\% of cluster galaxies exhibit recent merging features ($\mu_{r'}<27~\rm{mag}~\rm{arcsec^{-2}}$), while 4\% show ongoing merging features.
From a higher fraction of recent merging features, they concluded that the cluster environment is unsuitable for merging events between galaxies, and galaxies showing recent merging features experienced these events before falling into the cluster.
These observational studies have suggested that the details of the tidal features, such as morphological characteristics, prominence, and number, provide hints about the mass ratio of the merger, the orbital information, and the time since the last merging event, as tidal features are remnants of recent and/or ongoing mergers between galaxies.
Furthermore, as these features persist for long periods in the LSB regime, studies have shown that deeper imaging surveys can unveil many tidal features around satellite galaxies.
Given that the K-DRIFT can detect LSB structures down to 29-30$~\rm{mag}~\rm{arcsec^{-2}}$ in 10 hours of observation, many obscured tidal features around cluster galaxies are expected to be revealed, offering crucial insights to the hierarchical merging events of clusters.


{\bf Jellyfish galaxies:} Galaxies always experience ram pressure when moving through fluids. Galaxy clusters are the environments in which ram pressure is particularly influential on galaxies due to their high peculiar velocities and high intracluster medium density~\citep{1972ApJ...176....1G}. Numerous galaxies with tail-like features behind them have been observed in the HI, CO, H$\alpha$, and X-ray bands in cluster environments~\citep{gavazzi01,kenney04,wang04,finoguenov04,machacek05,oosterloo05,sun05,sun06,cortese06,cortese07,sun07,chung07,chung09,sun10,scott10,scott12,fumagalli14,boselli16,poggianti17,sheen17,scott18}. It has been suggested that ram pressure stripping (RPS) is the primary origin of the tail structures, and galaxies with characteristic features are often referred to as jellyfish galaxies~\citep{2009MNRAS.399.2221B}. Although RPS eventually quenches star formation activities by removing the ISM from galaxies~\citep{koopmann04a,koopmann04b}, this effect appears to cause a complicated evolution history for RPS galaxies on a short timescale. For instance, \cite{lee20} demonstrated that the mild ICM winds, mimicking ram pressure in the cluster outskirts, can enhance the star formation activities in RPS disks by compressing dense molecular clouds without producing clumpy structures in the tails. \citet{lee22} found that apparent jellyfish features can form when a gas-rich galaxy encounters strong ram pressure. In their simulations, an abundant ISM is stripped by ram pressure, mixing with the ICM and forming plenty of warm ionized clouds in the RPS tails that are bright in the H$\alpha$ band. The warm gas cools and collapses within a few hundred Myr across the RPS tails, eventually forming stars that are detected as H$\alpha$ knots in observations. \citet{lee22} also showed that H$\alpha$ emission closely correlates with the degree of mixing between the ISM and ICM in the RPS tails. While the jellyfish features are building up, disk star formation activities are quickly quenched by severe stripping. These results indicate that the impact of ram pressure varies significantly among galaxies with different gas fractions, masses, and orbits, even within a single cluster. 

Deep and wide-field observations are essential to explore the relationship between infall stages and the jellyfish features of cluster satellites. The unique optical design of K-DRIFT will enable us to find galaxies with various jellyfish features in cluster environments efficiently. For a more in-depth exploration of science cases related to the ICM, please refer to the corresponding paper in the K-DRIFT white paper series, Seon et al. (2025, in preparation).


{\bf Quenched fraction of satellite galaxies:} Many environmental processes, including those mentioned above, simultaneously affect cluster galaxies, leading to suppressed star formation activity \citep{2000ApJ...540..113B, 2004MNRAS.353..713K}.
Although many cluster galaxies show distinct characteristics compared to their field counterparts, it is still challenging to identify the primary mechanism responsible for this discrepancy.
However, the timescale over which star formation (or star formation rate) declines, "the quenching timescale," offers a useful diagnostic for determining the dominant process driving star formation suppression in cluster galaxies \citep{2013MNRAS.432..336W, 2018ApJ...866..136F, 2020ApJS..247...45R, 2021MNRAS.501.5073O, park2023}.
For example, if cluster galaxies have star formation quenching timescales comparable to their dynamical timescales within clusters, environmental processes related to their orbital motion (e.g., RPS) are likely to be the most significant.

The quenched fraction of cluster galaxies as a function of stellar mass has been widely used as a proxy for estimating the quenching timescale: low (high) quenched fractions indicate slow (fast) star formation quenching \citep[][]{2013MNRAS.432..336W, 2014MNRAS.438.3070M, 2016MNRAS.463.3083O}.
Recent large-scale observations have reported increasing quenched fractions with increasing stellar mass of galaxies.
Higher quenched fractions in massive galaxies could be interpreted as evidence for faster quenching, despite their stronger restoring force against cluster environmental effects.
However, this is unlikely to be accurate, because massive galaxies are typically subject to significant mass quenching before entering the cluster, resulting in low gas content at the time of infall \citep[e.g.,][]{2018ApJ...865..156J}.

Lower quenched fractions in low-mass galaxies, on the other hand, suggest longer quenching timescales, which appears counter-intuitive given their weaker restoring force \citep[][]{2014MNRAS.442.1396W, 2016MNRAS.463.3083O, 2020ApJS..247...45R}.
One possible explanation for this trend is observational selection bias due to the low completeness for low-mass galaxies; many quenched low-mass galaxies may simply remain undetected.
Indeed, the lower detection limit for cluster galaxies in recent large-scale surveys is around $10^{10}\,M_{\odot}$, making the determination of quenched fractions for galaxies with $M_{\star} < 10^{10}\,M_{\odot}$ less reliable.
Furthermore, recent cosmological simulations focusing on galaxy clusters fail to reproduce satellite galaxies below $M_{\star} \sim 10^{9-10}\,M_{\odot}$, mainly due to insufficient resolution to resolve them.
Therefore, there are still many opportunities to test the reliability of the quenched fraction at the low-mass end.

The deep and wide-field observations provided by K-DRIFT will significantly expand the galaxy sample in clusters to include the low-mass regime.
This will enable the determination of quenched fractions across a wide mass range, which has not been thoroughly explored yet, thereby offering new insights into the quenching mechanisms of cluster galaxies.

{\bf Hickson compact group:} \citet{hickson1982} reported one hundred compact groups (hereafter, HCGs) identified from the Palomar Observatory Sky Survey, including the well-known Seyfert's Sextet \citep{seyfert1948,seyferthst} and Stephan's Quintet \citep{stephan1877,stephanjwst}, which were listed as HCG 79 and HCG 92, respectively. 
The median redshift of the HCGs is $z = 0.030$, and most HCGs are associated with sparsely populated loose groups and filaments rather than dense rich clusters. 
However, the fraction of late-type galaxies in these groups is lower than that of field galaxies in low-density environments, and the contents of neutral gas in these compact groups are a factor of two lower than in isolated field galaxies, suggesting that HCGs are dynamically interacting systems \citep{hickson1997}. 

Narrow-band H$\alpha$ observations of 16 HCGs were performed by \citet{vilchez1998}. 
A significant fraction of elliptical and lenticular galaxies exhibited H$\alpha$ emissions. 
In three particular groups, H$\alpha$ emission was also detected in their tidal features, implying the potential debris of disrupted dwarf galaxies. 
Since K-DRIFT is designed for faint extended emissions with a wide FoV, broad- and/or narrow-band observations, centered at H$\alpha$, of HCGs can provide valuable insights into the interaction and evolution of compact galaxy groups within the context of galaxy formation and evolution.

\subsection{Large-Scale Structures}

Galaxy clusters typically reside at the intersections of cosmic filaments, which are prominent features of the large-scale structure of the universe. The prevailing model for structure formation draws inspiration from Zel'dovich's pancake concept \citep{1970A&A.....5...84Z}. This model posits that an initially inhomogeneous mass distribution experiences gravitational tidal forces, initiating a mildly non-linear stage of evolution. The process involves mass contraction along one axis, forming sheet-like walls, followed by filaments along the second axis, ultimately culminating in the collapse to form galaxy clusters.

A recent investigation has revealed a robust correlation between the ellipticity of more than 200 BCGs and their alignment with large-scale structures, extending out to ten times the cluster's characteristic radius, $R_{200}$ \citep{2023MNRAS.525.4685S}. This correlation suggests that the preferential inflow of galaxies along interconnected filaments plays a pivotal role in shaping the structure of a cluster's BCG. Furthermore, a recent weak-lensing study has detected the dark matter components of intracluster filaments in the Coma cluster field \citep{2024NatAs...8..377H}. This finding suggests that these intracluster filaments may represent the terminal segments of large-scale cosmic filaments, particularly those associated with individual clusters.

When combined with spectroscopic and weak-lensing data of galaxy clusters and their surrounding filaments, the forthcoming K-DRIFT LSB imaging data will serve as a valuable resource for studying the cosmic assembly history and dynamical evolution in their vicinity. In particular, deep imaging of the extended diffuse outskirts of BCGs, tidal features surrounding member galaxies, and the spatial distribution of ICL, together with their correlation with filaments, is expected to provide detailed insights into the formation and evolution of cosmic structures.

\subsection{Cosmology}
Galaxy clusters offer a unique window into cosmology. They are sensitive to several key cosmological parameters, such as the nature of dark matter and dark energy, as well as the large-scale structure of the universe. Their enormous mass amplifies subtle cosmological effects, making them ideal laboratories for testing models of cosmic evolution \citep{2011ARA&A..49..409A}.

In the realm of cluster cosmology, the potential of ICL as a cluster mass proxy, as discussed in Section \ref{subsec:DMtracer}, is of particular interest. Its suggested role in improving cluster-finding algorithms significantly increases its importance. Cosmological studies that rely on galaxy cluster abundance measurements emphasize the critical need for precise and accurate cluster mass proxies. Proxies with minimal scatter in relation to the true cluster mass can notably reduce the need for subsequent observations, thereby minimizing uncertainties in derived cosmological parameters like $\Omega_{M}$ and $\sigma_{8}$ \citep{2010ApJ...708..645R}.
The accuracy of such cosmological studies is contingent upon having a mass proxy that is robust against variations in the large-scale structural environment of the cluster, as underscored by \cite{2022MNRAS.515.4471W}. Incorporating diffuse light quantities in developing cluster mass proxies or cluster finding algorithms presents an intriguing avenue for further refinement, as proposed by \cite{2022MNRAS.515.4722H}.

{\bf $\Omega_{M}$ constraint:} Current measurements of the matter content energy budget of the universe ($\Omega_{M}$) heavily rely on observations of its bright components. However, traditional observations have overlooked a significant portion of the universe, particularly LSB structures such as ICL, UDGs, and dwarf galaxies. These LSB structures contribute a substantial fraction, approximately 15\%, of the total baryonic mass \citep{2021MNRAS.508.2634Y}. Moreover, recent studies have identified thousands of UDG candidates in the Coma cluster. By harnessing the capabilities of K-DRIFT to explore the LSB universe, we can significantly improve our understanding of the baryonic content and contribute to placing more accurate constraints on $\Omega_{M}$.


{\bf Test of modified gravity:} The recent discovery of the UDGs NGC 1052-DF2 and NGC 1052-DF4, which exhibit a velocity dispersion reasonably explained solely by the presence of baryonic matter without requiring any dark matter contribution, has opened up new avenues for exploring alternative theories of gravity \citep{2019MNRAS.482L...1M, 2023EPJC...83..402L}. This intriguing result motivates an investigation into modified gravity theories, such as f(R) gravity and the Renormalization Group correction to General Relativity. The line-of-sight velocity dispersion analysis of NGC 1052-DF2 and NGC 1052-DF4 indicates that certain f(R) gravity models, such as Taylor expanded f(R) about R = 0 or simple power-law models with of the form $f(R)\propto R^n$, are consistent with the observed data \citep{2023PhRvD.108f4021B}.

By including these recent findings on the velocity dispersion of dark-matter-deficient UDGs and their compatibility with alternative gravity theories, K-DRIFT's observations of LSB structures in galaxy clusters, along with the potential for a future space-based extension, may offer new insights into cosmological models, including those that extend beyond the standard $\Lambda$CDM framework.


\section{Strategies for LSB studies\label{sec:strategy}}

The exploration of LSB phenomena, as outlined in Section \ref{sec:science}, necessitates a comprehensive strategy that overcomes the limitations of current astronomical instrumentation. Although the importance of observing LSB structures such as ICL and UDG is widely recognized, actual observational studies remain scarce due to the inherent challenges of detecting such faint features \citep{2002ApJ...575..779F, 2004ApJ...617..879L, 2005ApJ...618..195G, 2005MNRAS.358..949Z, 2015ApJ...807L...2K, 2015ApJ...798L..45V, 2016ApJ...828L...6V, 2016ApJS..225...11Y, 2017ApJ...844L..11V, 2017ApJ...834...16M, 2018MNRAS.474.3009D, 2018ApJ...862...95K, 2018MNRAS.478.2034R, 2018ApJ...862...82L, 2018ApJ...859...37G, 2019A&A...622A.183J, 2020MNRAS.496.3182A, 2021MNRAS.508.2634Y, 2021MNRAS.503..679M, 2022MNRAS.512.3230M}. To effectively study these elusive structures, which are typically about 100 times fainter than the night sky, a paradigm shift in observational strategy is required. This involves not only collecting more photons to reduce Poisson noise but also optimizing telescope design to minimize systematic errors. In this context, smaller telescopes with simpler optics, contrary to the traditional approach of using large, complex structures, could prove more effective.

\subsection{Wide-Deep Survey}

The study of LSB phenomena, particularly in the context of galaxy clusters and ICL, requires wide-field and uniformly deep observations (exceeding 26.5 mag arcsec$^{-2}$). This necessity arises from the need to detect extended sources comparable in size to galaxy clusters and to accurately determine the level and pattern of the sky background accurately. Hence, minimizing systematic errors is crucial in pushing down the surface brightness limit and ensuring reliable measurements.

The measurement of ICL, in particular, requires a specialized observation strategy and meticulous data reduction \citep{2005ApJ...631L..41M, 2015MNRAS.446..120D, 2015A&A...581A..10C, 2016ApJ...823..123T, 2019A&A...621A.133B, 2020A&A...644A..42R, 2021ApJ...910...45M}. One of the critical challenges in this context is achieving an exceptionally low sky background error, which is pivotal for separating ICL components from the contributions of luminous objects such as stars and galaxies. The primary sources of sky background error are typically large-scale residuals arising from flat-fielding inaccuracies and difficulties in sky background subtraction.
To address these issues, K-DRIFT will implement several key strategies:

{\bf Large FoV with extensive dithering:} K-DRIFT’s FoV, approximately 4.5 $\times$ 4.5 deg$^2$, is optimally suited for detecting ICL across a wide area. The use of large dithering patterns in observations will be instrumental in avoiding issues related to imperfect flat-fielding and the over-subtraction of the sky. This approach ensures that the entire extent of the ICL, as well as the surrounding sky, is sampled accurately and uniformly.

{\bf Extended PSF wing modeling:} The extended wings of the PSF from bright stars can significantly contaminate LSB observations. K-DRIFT will employ advanced techniques to model these PSF wings accurately \citep{2025PASP..137e4502B,2025arXiv251022250K} and subtract their contribution from the data, thereby minimizing contamination in ICL measurements.

{\bf Uniform survey depth:} With consistently measured cluster light ICL can be rigorously identified and extracted. This process allows for the accurate calculation of ICL fraction, surface brightness profile, and color profile. The precision of these measurements depends critically on minimizing the sky background error, which will be a primary focus of K-DRIFT’s uniformly deep and wide survey approach.

Through these methodologies, K-DRIFT is poised to make significant contributions to the study of ICL and galaxy clusters. The telescope’s design and observation strategy are specifically tailored to overcome the challenges inherent in LSB studies, paving the way for groundbreaking discoveries and a deeper understanding of these elusive cosmic structures.




Our wide-deep survey will cover approximately 
1100 galaxy clusters, initially selected from 12 catalogs, including UPcluster-SZ \citep{2024ApJS..272....7B}, RedMaPPeR \citep{2016ApJS..224....1R}, eFEDS \citep{2022A&A...661A...2L}, eRASS \citep{2024A&A...682A..34M}, and DESI \citep{Zou_2021}. These clusters will then be filtered by redshift ($z < 0.1$) and declination ($-40^\circ<$ Dec $<-20^\circ$) to target the most observable systems. The survey may later be extended to cover a wider declination range (Dec $<0^\circ$) encompassing approximately 4300 galaxy clusters.

\subsection{Narrow-Band Imaging}
The identification of ICL as originating from intergalactic sources is complicated by the presence of central cD galaxies within many galaxy clusters. These galaxies exhibit extended surface brightness profiles that exceed the expected extrapolation of a de Vaucouleurs profile at large radii, making it difficult to distinguish the diffuse intracluster component from the light associated with the cD galaxy's halo.
One potential solution to this issue lies in kinematic observations of galaxy clusters, as demonstrated in previous studies through investigation of discrete tracers of the ICL, such as planetary nebulae, red giant stars, supernovae,  and globular clusters \citep[see][and references therein]{Arnaboldi22}. However, at present, general spectroscopic observations are not well suited to probing line emission from those individual tracers of the ICL (i.e., tracing the radial extent and the kinematics of the intracluster components across the entire galaxy cluster). Although high-performance integral field spectrometers are an option, their limited FoV, typically less than 1 arcmin$^2$, makes them inefficient for this purpose. An effective alternative approach to obtain low-resolution spectral maps down to faint surface brightness limits over very wide areas of the sky is narrow-band imaging \citep[e.g.,][]{Lokhorst19, Lokhorst20}.

{\bf [\ion{O}{iii}] 5007 \r{A} emission line:} Planetary nebulae (PNe) are often used to estimate the kinematics of diffuse stellar halos in galaxies and ICL, where analyzing individual stellar spectra is not feasible, because of their prominent [\ion{O}{iii}] 5007 \r{A} emission line \citep{Jacoby90, Arnaboldi96, Feldmeier03}. Additionally, since PNe represent the final stage of evolution for low- to intermediate-mass stars, they tend to be separately distributed similarly to the majority of stars \citep[e.g.,][]{Ciardullo89, Ennis23}. By comparing the spatial distribution of intracluster PNe with other components (e.g., galaxies, hot gas, and dark matter) of the galaxy cluster, we can obtain hints about the origin of the ICL \citep[][]{Arnaboldi10}. Furthermore, conducting detailed spectroscopic analyses (e.g., as a K-GMT large program) on intracluster PNe candidates identified through narrow-band imaging will help validate our ICL identification method based on comprehensive broad-band (\textit{u, g, r}) survey data.

The K-DRIFT survey of nearby galaxy clusters, using both broad- and narrow-band imaging, will yield significant findings in the field of ICL research, particularly through the analysis of intracluster PNe.

{\bf H$\alpha$ 6563 \r{A} emission line:} Observations of the circumgalactic medium (CGM) surrounding galaxies allow us to trace their extended environments and a variety of physical processes (e.g., infalling gas from the large scale environment, outflowing gas via active galactic nucleus (AGN) or starburst activity, and tidal stripping), that influences how the CGM affects the evolution of galaxies \citep[for reviews, see][]{2023ARA&A..61..131F, Crain23}. Although H$\alpha$ emission from the CGM is very faint \citep[e.g.,][]{Lokhorst20}, recent studies have shown that wide-field deep H$\alpha$ imaging with fast optical systems allows for the exploration of the diffuse emission from the extended CGM over tens of kpc \citep[e.g.,][]{Watkins17, Watkins18, Lokhorst22}.
Deep H$\alpha$ imaging with K-DRIFT has the potential to map the diffuse CGM surrounding galaxies in the nearby universe.

\subsection{NUV Imaging} 

The near-ultraviolet (NUV) is recognized as a sensitive proxy for star formation over the past $\sim$1 Gyr. As an indicator of recent star formation, NUV observations reveal a diversity among early-type galaxies, reflecting variations in the degree of recent star formation activity \citep{2000ApJ...541L..37F, 2005ApJ...619L.111Y, 2007ApJS..173..512S, 2007ApJS..173..619K, 2013ApJ...767...90K, 2016ApJ...820..132K}. Therefore, NUV observations of the ICL and other LSB features in galaxy clusters could serve as a valuable addition to the K-DRIFT science program, providing crucial insights into their origins. 

\begin{figure*}
    
\centering
\includegraphics[trim={0 10cm 0 0},clip,width=0.9\columnwidth]{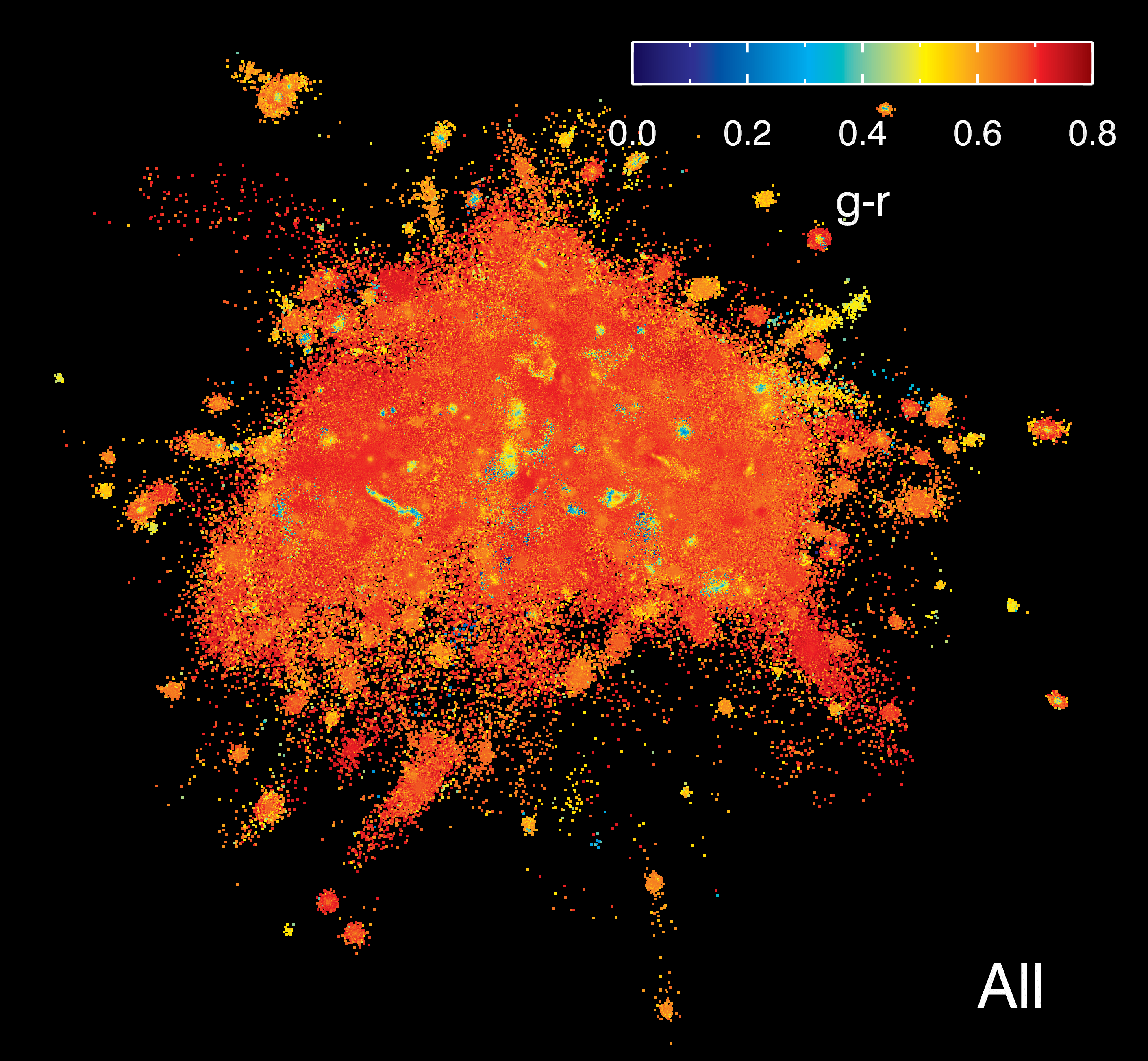}
\includegraphics[trim={0 10cm 0 0},clip,width=0.9\columnwidth]{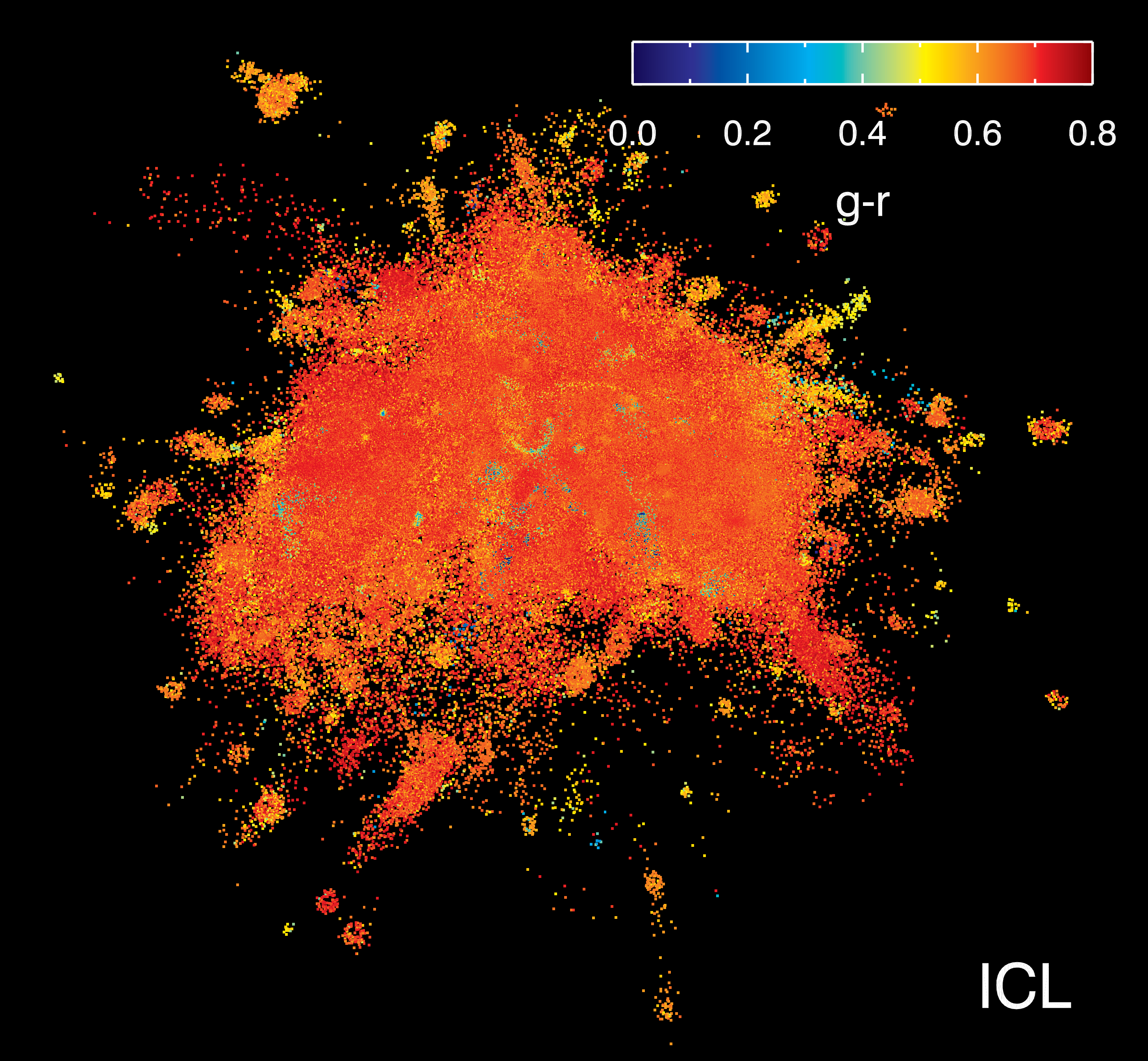}
\includegraphics[trim={0 10cm 0 0},clip,width=0.9\columnwidth]{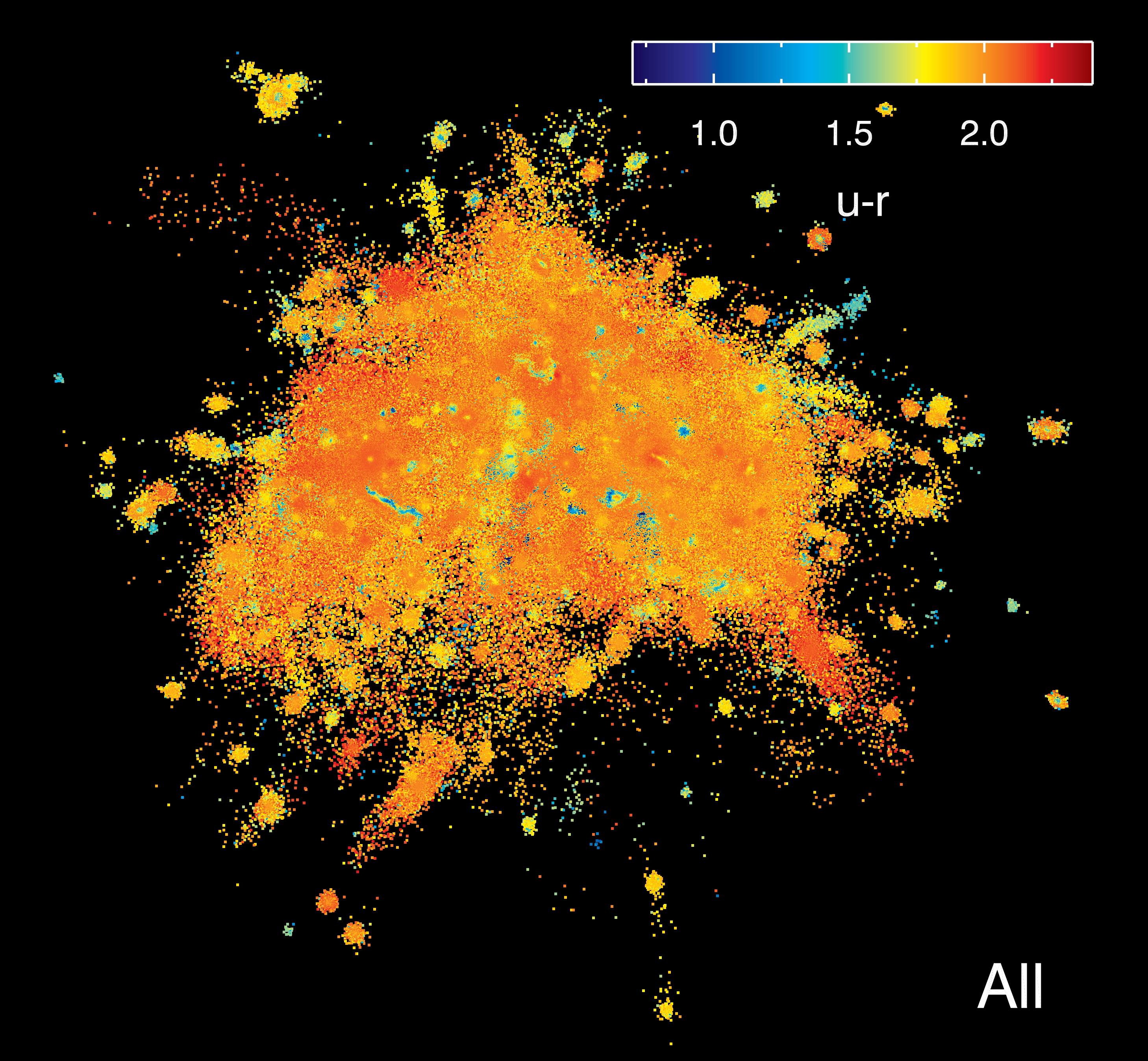}
\includegraphics[trim={0 10cm 0 0},clip,width=0.9\columnwidth]{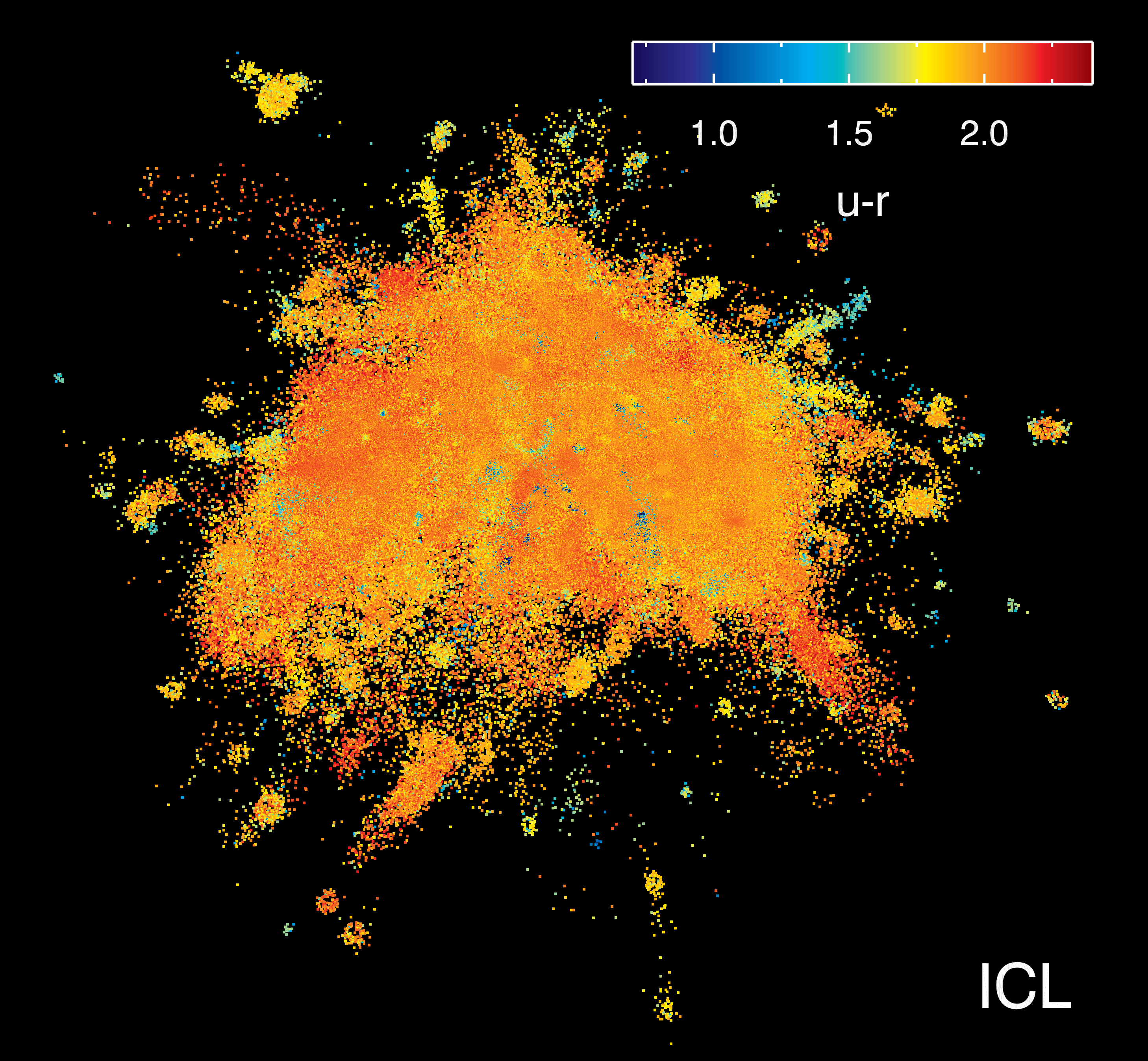}
\includegraphics[width=0.9\columnwidth]{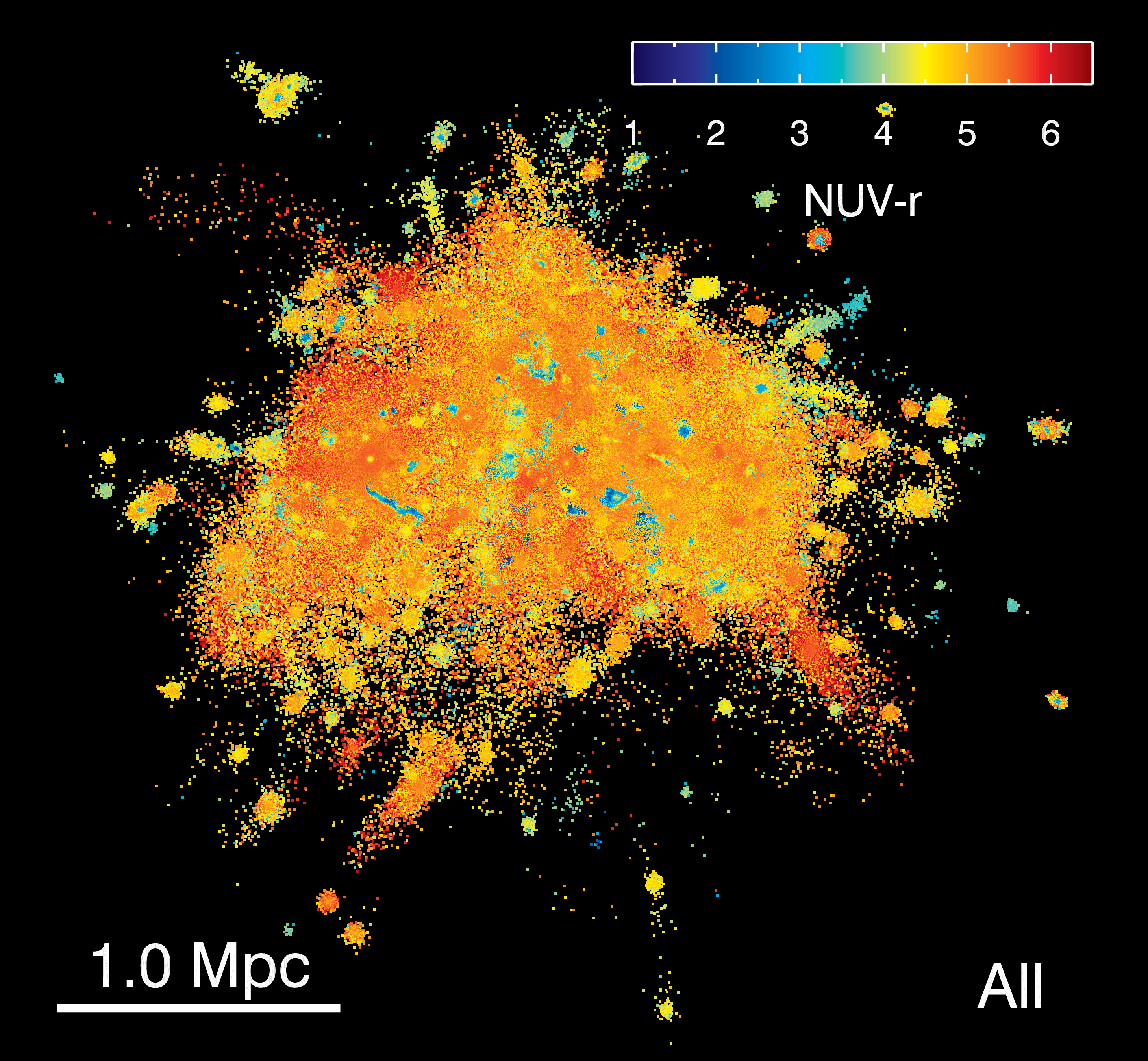}
\includegraphics[width=0.9\columnwidth]{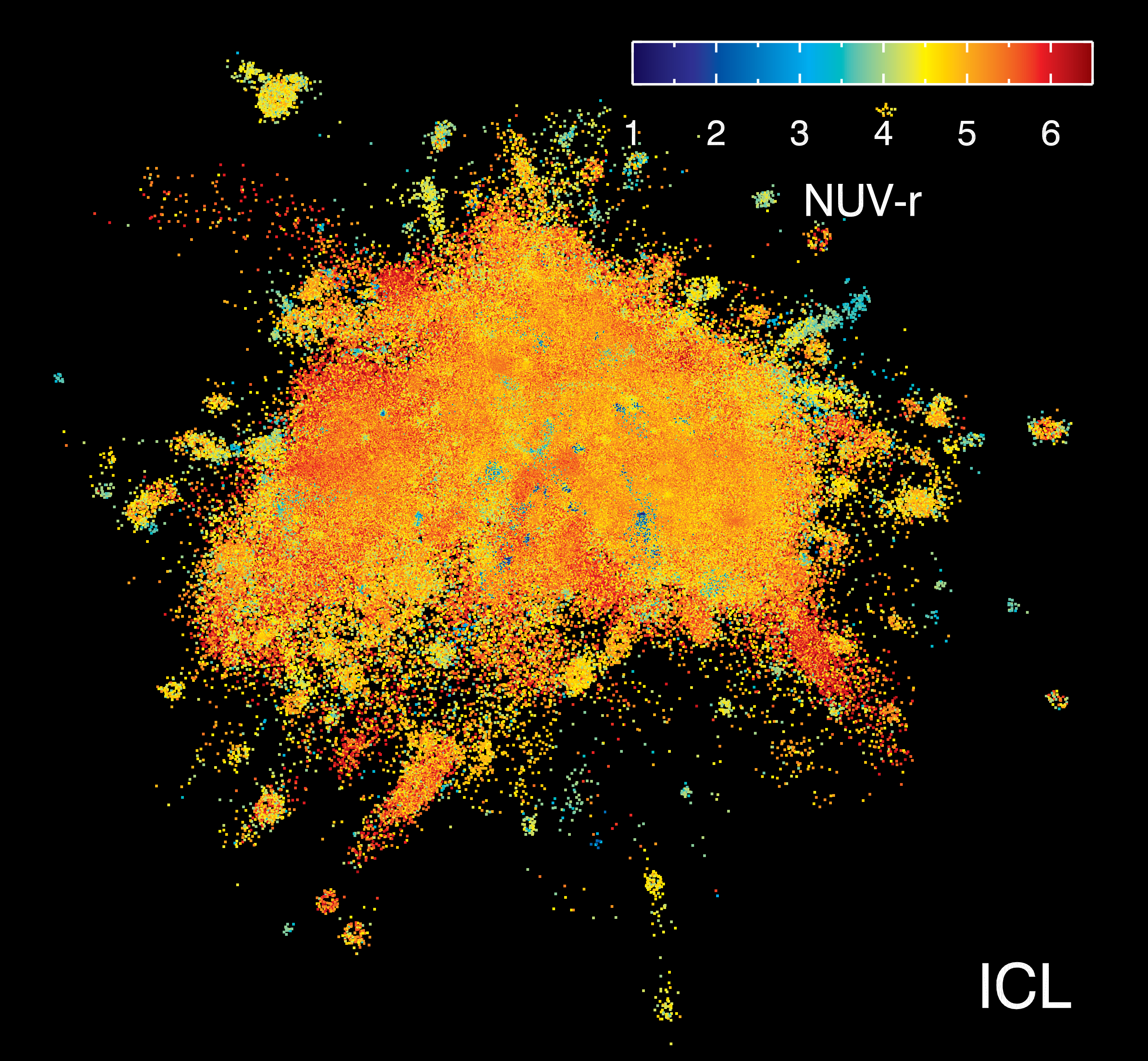}
\caption{
Mock color images of all cluster light (left) and ICL (right) from the Horizon Run 5 simulation. The top, middle, and bottom panels show the $g-r$, $u-r$, and NUV$-r$ mock color images, respectively. Theses mock color images were produced with a 2 arcsec pixel scale and a 1.5 arcsec PSF FWHM, simulating the anticipated performance of K-DRIFT. Notably, the NUV$-r$ color map exhibits greater variations compared to the $g-r$ and $u-r$ color maps.}
\label{fig:NUV}
\end{figure*}

To highlight the necessity of NUV observations with K-DRIFT, we generated mock images of a galaxy cluster from the HR5 simulation in $g-r$, $u-r$, and NUV$-r$ colors (see Figure \ref{fig:NUV}). The NUV, $u$, and $r$ magnitudes were computed using the mass-to-light ratios of a single stellar population in the three filter bands provided by the \texttt{E-MILES} stellar population synthesis model~\citep{vazdekis10,vazdekis16,ricciardelli12}. The galaxy cluster has a mass of $M_{\rm 200} \sim 4 \times 10^{14} M_{\odot}$ at the final redshift of the HR5 ($z=0.625$), and the stellar ages are adjusted assuming the cluster is located at $z=0.023$. A pixel scale of 2 arcsec and a PSF FWHM of 1.5 arcsec were adopted to create the mock color images, designed to mimic K-DRIFT's performance. Compared to the $g-r$ (upper panel) or $u-r$ (middle panel), the NUV$-r$ (lower panel) color image displays a broader color range in both the total cluster light (left) and ICL (right) maps. This broader color variation enables a more detailed investigation of the ICL's formation history compared to other color indices.

Building on the success of the ground-based K-DRIFT, we plan to develop its next-generation counterpart in space, dubbed the space-based K-DRIFT. With this space-based mission, we aim to explore the NUV universe in greater depth, opening up new opportunities to investigate the formation and evolution of galaxies, galaxy clusters, and LSB features with unprecedented precision.

\subsection{Cluster/Group Simulations for K-DRIFT’s LSB Observations}
Recent advances in observational technology have led to the detection of previously unseen LSB objects~\citep[e.g.,][]{2022ApJ...940L..19L,2022ApJ...940L..51M}.
This suggests that our knowledge of the evolution of individual galaxies and their environments may have been constrained by observational surface brightness limits \citep[see][]{2019MNRAS.485..796M}.
K-DRIFT is expected to overcome this limitation through deeper observations of large samples of galaxy groups and clusters.
To understand the clusters and their substructures observed by K-DRIFT, theoretical studies using high-resolution numerical simulations that realistically implement even LSB objects will also be essential.

{\bf Full-box simulations:} To include multiple clusters in a full-box cosmological simulation, a box with a side length of at least 100 $h^{-1}\,{\rm Mpc}$ is required.
In recent years, many simulations have been carried out in cubic boxes of comparable volume (e.g., \citealp[Horizon-AGN:][]{Dubois14}; \citealp[Illustris:][]{Vogelsberger14}; \citealp[EAGLE:][]{Schaye15}; \citealp[HR5:][]{lee21}), successfully reproducing key statistics observed in large-scale surveys.

Despite these successes, full-box numerical simulations are inherently forced to compromise between box size and spatial resolution due to limited computational resources.
For example, TNG300 \citep[][]{2019ComAC...6....2N} and HR5 include cluster samples of more than a hundred due to their very large box sizes, but their baryon mass resolution of $M_\star\sim10^6\,M_{\odot}$ is insufficient to reproduce the LSB regime 
 of $>$ 29 mag arcsec$^{-2}$ in the local universe~\citep[see][for further details]{kim23}.
Conversely, TNG50 achieves a force resolution of $\sim$300 pc in a (51.7 cMpc)$^3$ volume, and thus it barely contains cluster-scale structures in its cosmological domain.



{\bf Zoom-in simulations:} The zoom-in technique is a practical way to overcome the trade-off between resolution and box size. In this approach, the region of interest within the simulation box is computed at high resolution while the rest of the volume is set to a coarse resolution.
The {\textsc NewHorizon2} simulation \citep[][]{Yi24} serves as an example of this approach.
{\textsc NewHorizon2}, a twin simulation of {\textsc NewHorizon} \citep[][]{Dubois21}, focuses on a field region within a spherical volume of a $10\,{\rm Mpc}$ radius extracted from the full cosmological box of HorizonAGN.
This region is simulated with very high resolutions (e.g., $68\,{\rm pc}$ spatial resolution), including on-the-fly computations of the chemical evolution of nine elements (H, D, C, N, O, Mg, Si, S, and Fe).
This simulation allows for the study of galaxies down to $M_{\star} \sim 10^{7} M_{\odot}$ and hence provides a strong theoretical framework for understanding the formation and evolution of LSB objects with surface brightness as faint as $\sim$30 mag arcsec$^{-2}$ in field environments.
Furthermore, its counterpart, {\textsc NewCluster} \citep[][]{han2025} is a zoom-in simulation of a cluster region, with a resolution and physical models similar to those of {\textsc NewHorizon2}.
This simulation additionally includes on-the-fly calculations of the formation, coagulation, and destruction of four dust species, and is thereby expected to offer pioneering insights into the origin of LSB in cluster environments.
These simulations are anticipated to synergize well with the planned science programs with K-DRIFT, however, they are limited to a small number of clusters/groups, which is insufficient for conducting comprehensive statistical analyses.

{\bf GRT simulations:} Cosmological hydrodynamic simulations self-consistently follow the formation and evolution of galaxy clusters, modeling the hydrodynamics of the gas component from which stars form. However, such treatments of the baryonic component are highly computationally expensive. As a result, the spatial resolution of these simulations or the number of clusters that can be modeled is necessarily limited. To overcome the limitations of cosmological hydrodynamic simulations, \cite{chun2022} introduced an alternative simulation technique called the ``Galaxy Replacement Technique'' (GRT). The GRT is designed to follow the formation and evolution of clusters with a multi-resolution cosmological N-body simulation derived from the full merger tree of a low-resolution dark-matter-only cosmological simulation. This technique, based on a resimulation approach, allows for precise control of the properties and temporal evolution of the clusters while fully accounting for their cosmological context.

The GRT traces the spatial distribution and evolution of a cluster and its substructures without including computationally expensive baryonic physics. This efficient approach allows for studying large statistical samples of clusters with a high stellar mass resolution of $m_{\star} \sim 10^{5} M_{\odot}$ \citep{chun2023,chun2024}. 
Such high resolution enables us to accurately resolve tidal stripping processes and to describe in detail the formation of ultralow-surface-brightness features in clusters, such as the ICL and tidal features.

\cite{chun2024}, the most recent study using the GRT, performed simulations for 84 clusters with $13.6 < \log M_{\rm{200c}} [M_{\odot}] < 14.4$ to investigate the formation channels of the ICL within the clusters from $z=1.5$ to the present. The authors found that the ICL is already abundant at high redshift ($z=1.5$) and that its main formation channel is the tidal stripping of massive galaxies with stellar masses of $10 < \log M_{\star} [M_{\odot}] < 11$, regardless of redshift.
Additionally, they showed that the expected ICL fraction is strongly affected by observational detection limits, such as the observable radius determined by the telescope's FoV and the faint-end surface brightness limit (see Figure 12 in \citealp{chun2024}). In other words, the measured ICL fraction is changed by the choice of observable radius, and a higher surface brightness limit ($\mu_{V}<28.5~\rm{mag}~\rm{arcsec^{-2}}$) leads to an underestimation of the ICL fraction.
They also found that the relative importance of different ICL formation channels depends on the detection limit, with even the dominant channel changing under different observational conditions (see Figure 13 in \citealp{chun2024}).

\begin{figure*}
\centering
\includegraphics[width=0.9\textwidth]{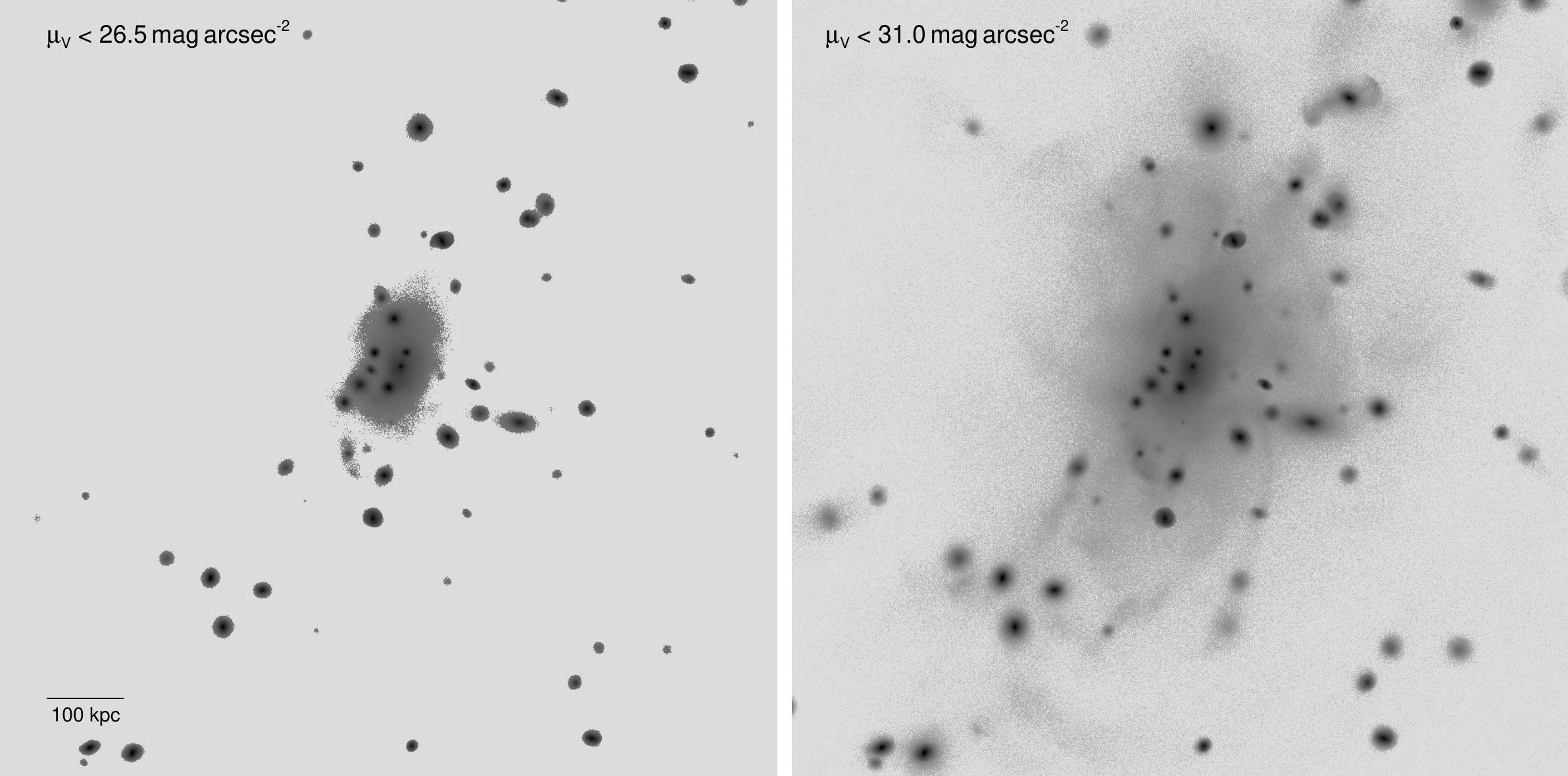}
\caption{Structures of a GRT cluster of $M_{\rm{200c}} = 2\times10^{14}$ $M_{\odot}$ brighter than $\mu_{V} <$ 26.5 mag arcsec$^{-2}$ (left) and  $\mu_{V} <$ 31 mag arcsec$^{-2}$ (right).
\label{fig:GRT1}}
\end{figure*}

Figure \ref{fig:GRT1} shows that not only the ICL but also the tidal features of the surviving satellites may be obscured by the detection limit.
\citet{adams2012} analyzed 54 clusters of 0.04 $<$ z $<$ 0.15, containing 3551 early-type galaxies, and identified tidal features. They found that $\sim$3\% of cluster early-type galaxies exhibit tidal features of $\mu_{r'}<26.5~\rm{mag}~\rm{arcsec^{-2}}$, and that the fraction of satellites with tidal features decreases at smaller clustercentric radii ($<$ 0.5 $R_{\rm{vir}}$).
On the other hand, in an ultralow-surface-brightness regime ($\mu_{V}\sim31.0~\rm{mag}~\rm{arcsec^{-2}}$), \cite{chun2025} show an increasing trend within the 84 GRT clusters.
This suggests that the numerous tidal features of satellites in the clusters might be veiled by detection limits.
The results of the GRT simulations emphasize the need for deeper imaging observations to investigate the ICL in greater detail.

While the GRT is an outstanding technique for tracing the formation and evolution of various LSB objects resulting from gravitationally merging events, its lack of hydrodynamics may cause it to miss the LSB structures originating from hydrodynamic processes.
To overcome this limitation, it is necessary to complement GRT-based studies with cosmological hydrodynamic simulations such as HR5, {\textsc NewCluster}, and IllustrisTNG.
Indeed, these simulation studies are expected to advance our understanding of LSB objects as well as clusters and their satellite galaxies within a comprehensive cosmological framework. 
Together, they are anticipated to contribute significantly to analysis of the clusters and their substructures observed by K-DRIFT.

\subsection{LSB Structure Detection using Machine Learning}


The LSB structures in galaxy clusters present substantial challenges for detection and quantitative analysis owing to their faint, diffuse emission and susceptibility to observational systematics. Recent advances in machine learning (ML), however, have demonstrated considerable potential for improving both the detection and characterization of such structures. For example, \citet{2021A&C....3500469T,2021OJAp....4E...3M} trained and evaluated a range of convolutional neural network (CNN) models on labeled LSB galaxies and imaging artifacts from the Dark Energy Survey, achieving classification accuracies exceeding 90\%. More recently, \citet{2024MNRAS.531.4070D} applied self-supervised representation learning to identify tidal features in Hyper Suprime-Cam imaging, while \citet{2023MNRAS.519.4735S} developed a U-net–based framework to automate the annotation of optical images containing Galactic cirrus emission. In \citet{2025_paudel, 2025A&A...701L...9P}, an external-attention model trained on a comprehensive catalog of dwarf elliptical (dE) galaxies \citep{2023ApJS..265...57P} was used to study the large-scale distribution of dEs in the Virgo Cluster, leading to the discovery of serendipitous systems such as the runaway galaxy NGC~524. Together, these studies illustrate the growing maturity of ML-based approaches for extracting scientifically valuable information from extremely low-contrast imaging data. The unique design of K-DRIFT, optimized specifically for LSB observations, makes it a particularly well-suited instrument for extending this line of research to galaxy clusters.

Machine learning algorithms can be trained to identify LSB features in K-DRIFT data, enabling efficient detection and extraction of relevant structures such as ICL, UDGs, and faint stellar streams. By training the algorithms on both simulated data and validated observations, these models can learn to recognize subtle signatures of LSB structures that are often difficult to detect using traditional analysis methods.

Additionally, machine learning techniques can aid in the classification and characterization of different types of LSB structures. For instance, convolutional neural networks can be utilized to classify UDG candidates based on their morphological properties, distinguishing them from other types of galaxies within the cluster environment. This automated classification enables large-scale studies of UDG populations and their relationship to the overall cluster properties.

Moreover, machine learning algorithms can facilitate the extraction of important physical parameters from LSB structures. By training models on synthetic data with known properties, these algorithms can estimate quantities such as the stellar masses, sizes, and luminosities of LSB objects in K-DRIFT observations. This information, combined with other observational data, enables detailed analyses of the formation, evolution, and environmental effects on LSB structures within galaxy clusters.

Integrating machine learning techniques into the analysis of K-DRIFT data not only enhances the detection and characterization of LSB structures but also improves the efficiency and scalability of scientific investigations. This synergy between K-DRIFT's unique observational capabilities and advances in machine learning opens up new avenues for studying the complex and intricate nature of LSB features in galaxy clusters, facilitating comprehensive analyses and deepening our understanding of the underlying astrophysical processes at play.

\section{Prospects for
Synergies with Other Surveys\label{sec:prospects}}
The K-DRIFT cluster survey research is poised to synergize with other ongoing and upcoming multi-wavelength surveys. In this section, we explore potential synergies and applications.

{\bf LSST:} The Vera C. Rubin Observatory is poised to generate a vast dataset through its extensive survey, the Legacy Survey of Space and Time (LSST), conducted in the southern hemisphere. LSST is expected to provide a catalog of approximately 20,000 clusters with mass estimates at about 10\% accuracy, along with 10,000 massive galaxy clusters extending to a redshift of approximately 1.2 \citep{2009arXiv0912.0201L, 2020arXiv200111067B, 2019ApJ...873..111I}.
The upcoming Wide-Fast-Deep survey of LSST promises a comprehensive photometric catalog spanning six broad-band filters ($ugrizy$; 320$-$1050 nm). Its wide FoV (9.6 deg$^2$) will be ideal for accurate sky background estimation, which is crucial for LSB studies. The expected surface brightness limit of LSST is $\mu_{r}^{\mathrm{limit}}$(3$\sigma$, $10'' \times 10''$) = 29.0 mag arcsec$^{-2}$ in its first year, and 30.3 mag arcsec$^{-2}$ after ten years of stacked data \citep{2018arXiv181204897L}. This advancement is expected to significantly improve masking performance of the diffuse outskirts of member galaxies, thereby strengthening the detection of LSB features such as the ICL. 
When combined with potential deep imaging from the NUV filter on the space-based K-DRIFT, this integrated approach could enable a more comprehensive exploration of recent star formation histories within galaxies residing in galaxy clusters.

{\bf DES:} The Dark Energy Survey \citep[DES;][]{2016MNRAS.460.1270D} is an astronomical survey conducted from 2013 to 2019 with the primary goal of constraining the properties of dark energy. It utilizes images captured in the NUV, visible, and near-infrared (NIR) wavelengths to measure the expansion of the universe through various methods, including Type Ia supernovae, baryon acoustic oscillations, galaxy clusters, and weak gravitational lensing.
The Dark Energy Camera (DECam), mounted on the 4-meter Víctor M. Blanco Telescope at the Cerro Tololo Inter-American Observatory in Chile, serves as the primary instrument for DES. DECam employs \textit{u, g, r, i, z,} and \textit{Y} filters covering a range of 340–1070 nm, similar to those used in the Sloan Digital Sky Survey. This enables DES to obtain photometric redshift measurements up to approximately $z \sim 1$. DECam features five lenses as corrector optics, extending the telescope's FoV to a diameter of 2.2 deg, one of the widest available for ground-based optical and infrared imaging, with a resolution of 0.263 arcsec.
Covering 5000 deg$^2$ of the southern sky, the DES survey area overlaps with the South Pole Telescope and Stripe 82, while strategically avoiding the Milky Way. The survey reaches a depth of 24th magnitude in the \textit{i}-band across the entire surveyed area, which is insufficient for the robust detection of tidal features. The high spatial resolution of DES imaging, when combined with the greater depth of K-DRIFT imaging, will provide a complementary dataset enabling comprehensive LSB studies across large samples of galaxy clusters.

{\bf SPHEREx:} The Spectro-Photometer for the History of the Universe, Epoch of Reionization and Ices Explorer \citep[SPHEREx;][]{2014arXiv1412.4872D} is a NIR space telescope currently conducting an all-sky spectral survey. Its comprehensive spectro-photometry maps encompass 102 spectral channels ranging from 0.75 to 5 $\mu$m, with a spectral resolution of 6 arcsec per pixel. SPHEREx observations will cover a vast array of celestial objects, including billions of galaxies and approximately 25,000 galaxy clusters.
Traditionally, LSB studies have been limited by sparse spectral data resulting from low signal-to-noise ratios and the high cost of observations. The ongoing SPHEREx spectro-photometry survey, which will provide a dataset equivalent to integral field unit observations across the entire sky, is poised to revolutionize our understanding of the kinematics and stellar populations of LSB objects. The synergistic combination of low-resolution (6 arcsec) spectrophotometric maps from SPHEREx and higher-resolution (2 arcsec) deep photometric maps from K-DRIFT is anticipated to yield groundbreaking insights into LSB studies.
By leveraging potential NUV data spanning 200–300 nm from the proposed space-based K-DRIFT mission, in combination with SPHEREx observations extending to 800 nm, we aim to construct the full spectral energy distributions of galaxies across thousands of galaxy clusters.

{\bf Euclid:} Euclid \citep{2025A&A...697A...1E}, a mission led by the European Space Agency, stands as a visible-to-NIR space telescope designed to unravel the complexities of the evolving dark universe. Its mission is to create a three-dimensional map of the universe, incorporating time as the third dimension, achieved through the measurement of shapes and redshifts of galaxies and clusters out to redshifts about 2.
Euclid is equipped with two scientific instruments: the visible-wavelength camera (VISible instrument), which covers a broad band from 550 to 900 nm with a spatial resolution of 0.1 arcsec, and the NIR camera/spectrometer (Near-Infrared Spectrometer and Photometer), which provides Y, J, and H broad-bands photometry and offers spectroscopy in the range of 1100 to 2000 nm with a spatial resolution of 0.3 arcsec and a spectral resolution of $R=250$.
Space-based surveys like Euclid provide invaluable imaging data for LSB studies, as they are significantly less affected by sky background interference compared to ground-based telescopes.
Euclid will conduct a $\sim$15,000 deg$^{2}$ weak-lensing survey, measuring the shapes of $\sim$1.5 billion galaxies. The resulting dark matter maps will provide a powerful reference for comparisons with the ICL distributions identified by K-DRIFT. In addition, Euclid’s imaging of large numbers of UDGs in three NIR filters will significantly enhance the interpretation of UDG populations discovered with K-DRIFT.

{\bf GALEX:} The Galaxy Evolution Explorer \citep[GALEX;][]{2007ApJS..173..682M} is a space telescope that observed galaxies in ultraviolet light, tracing 10 billion years of cosmic history. Launched in 2003 and operational for ten years until decommissioned in 2013, GALEX surveyed wide and deep fields in the NUV (175–280 nm) and far-UV (135–174 nm) wavelengths, with a 1.2 deg FoV and a spatial resolution of 5 arcsec. Leveraging the heritage from GALEX, in combination with deep imaging from K-DRIFT, will enable detailed exploration of the star formation in the distant universe.

{\bf Follow-up observation with large telescopes:} The K-GMT Science Program provides the Korean Astronomical Society with observational facilities to strength the community's scientific capabilities in the field of observational astronomy and astrophysics in preparation for the coming era of the Giant Magellan Telescope (GMT). Currently, the K-GMT Science Program provides observing opportunities at the Gemini observatory. The K-DRIFT cluster survey would provide excellent initial targets, including UDG, for conducting detailed spectroscopic and deep imaging observations using Gemini and GMT for more distant targets.

{\bf K-SPEC:} In order to investigate various properties of cluster galaxies and their hosts, it is important to distinguish bona fide member galaxies from other field galaxies. This requires accurate redshift information for individual galaxies, as it is the only way to measure the distances and relative velocities of distant objects. KASI-Spectrograph (K-SPEC) is a multi-object spectrograph that is under development by Korean researchers. A spectroscopic survey aimed to achieve the spectroscopic completeness of >90\% for galaxies with $Ks<13.75$ will be performed with K-SPEC. This survey is helpful in mapping the large-scale structure and identifying galaxy clusters and groups in the low-redshift universe, where K-DRIFT plans to conduct observations. The spectroscopic map from the survey will provide environment information for K-DRIFT galaxies, such as the redshifts and masses of their host clusters and groups, as well as the spectroscopic information for host galaxies showing LSB features. Therefore, by combining both survey datasets, we can expect to study the environmental effect on galaxy evolution and the spatial distribution of cluster galaxies.

{\bf uGMRT HI survey:} 
The galaxy cluster environment affects galaxy properties through both gravitational and hydrodynamic processes. One of the fundamental tools for understanding those effects is to study cluster galaxies' gas properties. 
Dynamical interactions, such as tidal stripping and in-situ star formation from stripped cold gas, contribute to the enrichment of ICL. Detection of extended structures of HI gas around galaxies in clusters, with morphologies aligned with the diffuse light distribution, could provide evidence for potential in-situ star formation.
The upgraded Giant Metrewave Radio Telescope (uGMRT) addresses these limitations and can probe star formation outside galaxies.
Focusing on the infalling group around NGC 4839 in the Coma cluster, which has been poorly covered by previous HI surveys, provides an opportunity to examine the in-situ ICL and intragroup light formation mechanism. Applying the WOC method to compare the spatial distribution of diffuse light and HI gas in this group could reveal valuable insights into in-situ star formation. The uGMRT observations on this subject were conducted in April-May 2023 (GTAC code: 44\_046, PI: Jaewon Yoo), with the potential to expand the survey based on the results.
The synergy between uGMRT HI survey data and future K-DRIFT data for galaxy clusters promises a lasting legacy for future studies.

{\bf SKA:}   
The Square Kilometre Array (SKA) telescopes are revolutionary instruments for radio astronomy, consisting of two world-leading, complementary arrays located on different continents. SKA-Mid, with 197 dish antennas, is being constructed in South Africa, while SKA-Low, featuring 131,072 tree-like antennas, is under construction in Western Australia. These arrays will span baselines of up to 150 km in South Africa and 74 km in Australia.
With their unparalleled sensitivity, the SKA telescopes will map hydrogen reservoirs around galaxies, the raw material for star formation, shedding light on how gas is accreted onto galaxies and converted into stars. When combined with future K-DRIFT observations of galaxy clusters, the SKA data will provide a deeper understanding of star formation histories and the evolution of galaxies and galaxy clusters.

\section{Conclusions\label{sec:conclusions}}
We explored the possible science cases related to galaxy clusters, which will be investigated using our new instrument developed for LSB science, K-DRIFT. K-DRIFT's specialized design positions it as a prime tool for probing the LSB features, such as ICL, faint substructures in BCG, UDG, and tidal features.
K-DRIFT will give hints about the key questions on the LSB field that lead to a step change in our knowledge of nearby galaxy clusters and groups.

ICL is known to encapsulate the dynamical interaction history of the encompassing galaxy cluster.
Leveraging K-DRIFT's sensitivity to faint signals allows for unprecedented studies on surface brightness and color profiles of ICL, which offer a unique vantage point for investigating its origin and the evolutionary processes of the host galaxy cluster. 
Moreover, ICL has recently emerged as a promising candidate for tracing dark matter. 
The telescope's ability to capture detailed ICL distributions will shed light on its potential role as a proxy for the elusive dark matter content within galaxy clusters. The investigation into the spatial distributions of ICL and dark matter, particularly in clusters with diverse dynamical states, promises valuable insights into cluster dynamics and the co-evolution of ICL with the host cluster.

K-DRIFT's advanced imaging capabilities extend beyond ICL studies, delving into the tidal streams and faint stellar halos, such as BCG inner structure and UDGs within galaxy clusters. This enables a comprehensive exploration of the underlying stellar content, offering valuable clues about the assembly history and interactions shaping these LSB objects.
Beyond the realm of galaxy clusters, K-DRIFT opens avenues for discovering and characterizing various LSB structures within the cosmic web, i.e., filaments and voids. The telescope's exceptional sensitivity to LSB allows for identifying dwarf galaxies, faint stellar streams, and other elusive astronomical entities in large-scale structures, contributing to a deeper understanding of cosmic structures.

We present our strategies for addressing the scientific inquiries introduced earlier. To minimize systematic errors, we propose a comprehensive, wide-deep survey. The consistent measurement of diffuse light across numerous galaxy groups and clusters in the southern hemisphere is poised to address a significant portion of questions in the LSB research field. Leveraging our LSB-specialized simulations and machine-learning techniques will be instrumental in interpreting observational results and refining our observation strategy.  Furthermore, we explore potential synergies between K-DRIFT and other ongoing or upcoming multi-wavelength surveys.

K-DRIFT stands at the forefront of LSB studies, offering unparalleled capabilities for unraveling the intricacies of galaxy clusters and exploring the cosmos at its faintest luminosities.


\acknowledgments
J.Y. was supported by a KIAS Individual Grant (QP089902) via the Quantum Universe Center at Korea Institute for Advanced Study.
K.C. was supported by the National Research Foundation of Korea (NRF) grant funded by the Korea government (MSIT) (2021R1F1A1045622). 
J.S. acknowledges support from the National Research Foundation of Korea grants (No. RS2025‐00516904 and No. RS‐2022‐NR068800) funded by the Ministry of Science, ICT \& Future Planning. 
C.G.S is supported via the Basic Science Research Program from the National Research Foundation of Korea (NRF) funded by the Ministry of Education (No. 2018R1A6A1A06024977).
J.L. acknowledges the support of the National Research Foundation of Korea (RS-2021-NR061998). 
J.R. was supported by the KASI-Yonsei Postdoctoral Fellowship and was supported by the KASI under the R\&D program (Project No. 2023-1-830-00), supervised by the Ministry of Science and ICT. 
H.K. acknowledges the support of the Agencia Nacional de Investigación y Desarrollo (ANID) ALMA grant funded by the Chilean government, ANID-ALMA-31230016.
This study was also supported by the National Research Foundation of Korea (NRF) grant funded by the Korea government (MSIT, 2022M3K3A1093827).
This work benefited from the outstanding support provided by the KISTI National Supercomputing Center and its Nurion Supercomputer through the Grand Challenge Program (KSC-2018-CHA-0003, KSC-2019-CHA-0002, KSC‐2022‐CRE‐0123). 
Large data transfer was supported by KREONET, which is managed and operated by KISTI. 
This work is also supported by the Center for Advanced Computation at Korea Institute for Advanced Study. 



 \bibliography{main}





\end{document}